%% file: manuscript.tex
\documentclass[12pt,twoside]{article}
\usepackage{latexsym}
\usepackage{amsfonts}
\usepackage{amssymb}
\usepackage{amsmath}
\usepackage{amstext}
\usepackage{makeidx}
\usepackage{graphicx}
\usepackage{bm}
\usepackage{times}
\makeindex
\usepackage{vmargin}
\setpapersize{A4}
\setmargins{2.6cm}{1.6cm}{16.0cm}{24.0cm}{15pt}{22pt}{0pt}{30pt} 
\usepackage{fancyhdr}
\newcommand{\MakeTitle}{
 \thispagestyle{empty}
 \renewcommand{\baselinestretch}{1.15}
 \Large\sf
 \begin{tabular}{p{0.0\linewidth}p{0.9\linewidth}}
  \LARGE\bf\TheChapter
   &{\raggedright\LARGE\bf\TheTitle    }\\[3ex]
   &{\raggedright\Large\sf\TheAuthor   }\\
   &{\raggedright\Large\sf\TheInstitute}\\
   &{\raggedright\Large\sf\TheAddress  }\\[16ex]
 \end{tabular} 
 \normalsize\rm
 \renewcommand{\baselinestretch}{1.0}
}

\usepackage{curves}
\usepackage{color} 
\usepackage{makeidx}
\newcommand{\mat}[1]{\mathbf{#1}}
\makeindex

\def\TheTitle{Theory and Practice of Density-Functional Theory}
\def\TheHeading{Theory and Practice of Density-Functional Theory} 
\def\TheAuthor{Peter E. Bl\"ochl}                     
\def\TheInstitute{Institute for Theoretical Physics}  
\def\TheAddress{Clausthal University of Technology}   
\def\TheChapter{}                                    

\begin{document}
\MakeTitle           
\tableofcontents     
\newpage

\section{Introduction}

On the nanoscale, materials around us have a surprisingly simple
structure: The standard model of solid state physics and chemistry
only knows of two types of particles, namely the nuclei making up the
periodic table and the electrons. Only one kind of interaction between
them needs to be considered, namely the electrostatic interaction.
Even magnetic forces are important only in rare occasions. All other
fundamental particles and interactions are irrelevant for chemistry.

The behavior of these particles can be described by the
Schr\"odinger equation (or better the relativistic Dirac equation),
which is easily written down. However, the attempt to solve this
equation for any system of interest fails miserably due to what Walter
Kohn termed the exponential wall\index{exponential wall}
\cite{kohn99_nobel}.

To obtain an impression of the powers of the exponential wall, imagine
the wave function of a N$_2$ molecule, having two nuclei and fourteen
electrons.  For $N$ particles, the Schr\"odinger equation is a partial
differential equation in $3N$ dimensions.  Let us express the wave
function on a grid with about 100 points along each spatial direction
and let us consider two spin states for each electron. Such a wave
function is represented by $2^{14}100^{3*16}\approx10^{100}$ complex
numbers. A data server for this amount of data, made of current
terabyte hard disks, would occupy a volume with a diameter of
$10^{10}$ light years!

Treating the nuclei as classical particles turned out to be a good
approximation, but the quantum nature of the electrons cannot be
ignored. A great simplification is to describe electrons as
non-interacting quasi particles. Instead of one wave function in $3N$
dimensions, one only needs to describe $N$ wave functions in three
dimensions each, a dramatic simplification from $10^{100}$ to $10^7$
numbers.

While the independent-particle model is very intuitive, and while it
forms the basis of most text books on solid-state physics, materials
physics, and chemistry, the Coulomb interaction between electrons is
clearly not negligible.

Here, density-functional
theory \cite{hohenberg64_pr136_B864,kohn65_pr140_1133} comes to our
rescue: it provides a rigorous mapping from interacting electrons onto
a system of non-interacting electrons.  Unfortunately, the exact
mapping is utterly complicated and this is where all the complexity
goes. Luckily, there are simple approximations that are both,
intuitive and surprisingly accurate. Furthermore, with the help of
clever algorithms, density-functional calculations can be performed on
current computers for large systems with several hundred atoms in a
unit cell or a molecule.  The microscopic insight gained from density
functional calculations is a major source of progress in solid
state physics, chemistry, material science, and biology.

In the first part of this article, I will try to familiarize the
novice reader with the basics of density-functional theory, provide
some guidance into common approximations and give an idea of the type
of problems that can be studied with density functional theory.

Beyond this article, I recommend the insightful review articles on
density functional theory by Jones and Gunnarsson
\cite{jones89_rmp61_689}, Baerends \cite{baerends97_jpca101_5383}, von
Barth \cite{vonbarth04_physicascripta109_9}, Perdew
\cite{perdew05_jcp123_62201}, Yang \cite{cohen08_science321_792}, and
their collaborators.

Solving the one-particle Schr\"odinger equation, which results from
density-functional theory, for real materials is a considerable
challenge. Several avenues have been developed to their solution. This
is the field of electronic structure methods, which will be discussed
in the second part of this article. This part is taken from earlier
versions by Clemens F\"orst, Johannes K\"astner and myself
\cite{bloechl05_book1_93, bloechl03_bms26_33}.

\section{Basics of  density-functional theory}
The dynamics of the electron wave function is governed by the
Schr\"odinger equation
$i\hbar\partial_t|\Psi\rangle=\hat{H}|\Psi\rangle$ with the N-particle
Hamiltonian $\hat{H}$.
\begin{eqnarray}
  \hat{H}=\sum_{j=1}^{N}
  \biggl(\frac{-\hbar^2}{2m_e}\vec{\nabla}_j^2+v_{ext}(\vec{r}_j)\biggr)
  +\frac{1}{2}\sum_{i\ne j}^N\frac{e^2}{4\pi\epsilon_0|\vec{r}_i-\vec{r}_j|}
\;.
\end{eqnarray}
With $m_e$ we denote the electron mass, with $\epsilon_0$ the vacuum
permittivity, $e$ is the elementary charge and $\hbar$ is the Planck
quantum divided by $2\pi$. The Coulomb potentials of the nuclei have
been combined into an external potential $v_{ext}(\vec{r})$.

All N-electron wave functions $\Psi(\vec{x}_1,\ldots,\vec{x}_{N})$
obey the Pauli principle, that is they change their sign, when two of
its particle coordinates are exchanged.

We use a notation that combines the position vector
$\vec{r}\in\mathbb{R}^3$ of an electron with its discrete spin coordinate
$\sigma\in\{\uparrow,\downarrow\}$ into a single vector
$\vec{x}:=(\vec{r},\sigma)$. Similarly, we use the notation of a
four-dimensional integral $\int d^4x:=\sum_{\sigma}\int d^3r$ for the
sum over spin indices and the integral over the position. With the
generalized symbol
$\delta(\vec{x}-\vec{x'}):=\delta_{\sigma,\sigma'}\delta(\vec{r}-\vec{r'})$
we denote the product of Kronecker delta of the spin coordinates and
Dirac's delta function for the positions.  While, at first sight, it
seems awkward to combine continuous and discrete numbers, this
notation is less error prone than the notation that treats the spin
coordinates as indices, where they can be confused with quantum
numbers. During the first reading, the novice will ignore the
complexity of the spin coordinates, treating $\vec{x}$ like a
coordinate. During careful study, he will nevertheless have the
complete and concise expressions.

\subsection*{One-particle reduced density matrix  and two-particle density}
In order to obtain the ground state energy
$E=\langle\Psi|\hat{H}|\Psi\rangle$ we need to perform $2^N$
integrations in $3N$ dimensions each, i.e.
\begin{eqnarray}
E&=&\int d^4x_1\cdots\int d^4x_N\;
\Psi^*(\vec{x}_1,\ldots,\vec{x}_N)\hat{H}\Psi(\vec{x}_1,\ldots,\vec{x}_N)\;.
\label{eq:etotwithwavef}
\end{eqnarray}

However, only two different types of integrals occur in the expression
for the energy, so that most of these integrations can be performed
beforehand leading to two quantities of physical significance.
\begin{itemize}
\item One of these quantities is the one-particle reduced density
  matrix\index{density matrix!one-particle}
  $\rho^{(1)}(\vec{x},\vec{x'})$, which allows one to evaluate all
  expectation values of one-particle operators such as the kinetic
  energy and the external potential,
\begin{eqnarray}
  \rho^{(1)}(\vec{x},\vec{x'})&:=&
  N\int d^4x_2\ldots\int d^4x_N\;
  \Psi(\vec{x},\vec{x}_2,\ldots,\vec{x}_N)
  \Psi^*(\vec{x'},\vec{x}_2,\ldots,\vec{x}_N)\;.
\end{eqnarray}
\item The other one is the two-particle
  density\index{density!two-particle} $n^{(2)}(\vec{r},\vec{r'})$,
  which allows to determine the interaction between the electrons,
\begin{eqnarray}
  n^{(2)}(\vec{r},\vec{r'})&:=&
  N(N-1)\sum_{\sigma,\sigma'}\int d^4x_3\ldots\int d^4x_N\;
  |\Psi(\vec{x},\vec{x'},\vec{x}_3,\ldots,\vec{x}_N)|^2\;.
\end{eqnarray}
\end{itemize}
If it is confusing that there are two different quantities depending
on two particle coordinates, note that the one-particle reduced
density matrix $\rho^{(1)}$ depends on two $\vec{x}$-arguments of the
same particle, while the two-particle density $n^{(2)}$ depends on the
positions of two different particles.

With these quantities the total energy is
\begin{eqnarray}
E&=&
\int d^4x'\int d^4x\;
\delta(\vec{x'}-\vec{x})
\left(\frac{-\hbar^2}{2m_e}\vec{\nabla}^2
+v_{ext}(\vec{r})\right)\rho^{(1)}(\vec{x},\vec{x'})
\nonumber\\
&&
+\frac{1}{2}\int d^3r\int d^3r'\;
 \frac{e^2n^{(2)}(\vec{r},\vec{r'})}{4\pi\epsilon_0|\vec{r}-\vec{r'}|}
\;,
\label{eq:etotwithrho1andn2}
\end{eqnarray}
where the gradient of the kinetic energy operates on the first argument
$\vec{r}$ of the density matrix.

\subsection*{One-particle reduced density matrix and natural orbitals}
In order to make oneself familiar with the one-particle reduced
density matrix, it is convenient to diagonalize it. The eigenstates
$\varphi_n(\vec{r})$ are called natural orbitals\index{natural
  orbital} \cite{loewdin55_pr97_1474} and the eigenvalues $\bar{f}_n$
are their occupations\index{occupation}. The index $n$ labels the
natural orbitals may stand for a set of quantum numbers.

The density matrix can be written in
the form
\begin{eqnarray}
\rho^{(1)}(\vec{x},\vec{x'})=\sum_n \bar{f}_n 
\varphi_n(\vec{x})\varphi^*_n(\vec{x'})
\;.
\end{eqnarray}
The natural orbitals are orthonormal one-particle orbitals, i.e.
\begin{eqnarray}
\int d^4x\;\varphi^*_m(\vec{x})\varphi_n(\vec{x})=\delta_{m,n}
\;.
\end{eqnarray}

Due to the Pauli principle, occupations are non-negative and never
larger than one \cite{coleman63_rmp35_668}. The natural orbitals
already point the way to the world of effectively non-interacting
electrons.

The one-particle density matrix provides us with the electron density 
\begin{eqnarray}
n^{(1)}(\vec{r})=\sum_\sigma \rho^{(1)}(\vec{x},\vec{x})
=\sum_\sigma\sum_{n} \bar{f}_n \varphi_n^*(\vec{x})\varphi_n(\vec{x})
\;.
\end{eqnarray}

With the natural orbitals, the 
total energy Eq.~\ref{eq:etotwithrho1andn2} obtains the form
\begin{eqnarray}
E=\sum_n \bar{f}_n \int d^4x\; 
\varphi_n^*(\vec{x})\frac{-\hbar^2}{2m}\vec{\nabla}^2\varphi_n(\vec{x})
&+&\int d^3r\; v_{ext}(\vec{r})n^{(1)}(\vec{r})
\nonumber\\
&+&
\frac{1}{2}\int d^3r\int d^3r'\;
\frac{e^2n^{(2)}(\vec{r},\vec{r'})}
{4\pi\epsilon_0|\vec{r}-\vec{r'}|}\;.
\label{eq:etotwithnatorbandn2}
\end{eqnarray}

\subsection*{Two-particle density and exchange-correlation hole}
The physical meaning of the two-particle density
$n^{(2)}(\vec{r},\vec{r'})$ is the following: For particles that are
completely uncorrelated, meaning that they do not even experience the
Pauli principle, the two particle density would be\footnote{This is
  correct only up to a term that vanishes in the limit of infinite
  particle number.} the product of one-particle densities,
i.e. $n^{(2)}(\vec{r},\vec{r'})=n^{(1)}(\vec{r})n^{(1)}(\vec{r'})$.
If one particle is at position $\vec{r}_0$, the density of the
remaining $N-1$ particles is the conditional
density\index{density!conditional}
\begin{eqnarray*}
\frac{n^{(2)}(\vec{r}_0,\vec{r})}{n^{(1)}(\vec{r}_0)}\;.
\end{eqnarray*}
The conditional density is the electron density seen by one of the
electrons at $\vec{r}_0$. This observer electron obviously only sees
the remaining $N-1$ electrons.

It is convenient to express the two-particle density by the hole
function\index{hole function} $h(\vec{r},\vec{r'})$, i.e.
\begin{eqnarray}
n^{(2)}(\vec{r},\vec{r'})
=n^{(1)}(\vec{r})\biggl[n^{(1)}(\vec{r'})+h(\vec{r},\vec{r'})\biggr]\;.
\label{eq:divisionhole}
\end{eqnarray}
One electron at position $\vec{r}$ does not ``see'' the total electron
density $n^{(1)}$ with $N$ electrons, but only the density of the
$N-1$ other electrons, because it does not see itself. The hole
function $h(\vec{r_0},\vec{r})$ is simply the difference of the total
electron density and the electron density seen by the observer
electron at $\vec{r}_0$.

The division of the two-particle density in Eq.~\ref{eq:divisionhole}
suggests to split the electron-electron interaction into the so-called
Hartree energy\index{Hartree energy}
\begin{eqnarray}
E_H\stackrel{\text{def}}{=}\frac{1}{2}\int d^3r\int d^3r'\;
\frac{e^2n^{(1)}(\vec{r})n^{(1)}(\vec{r'})}
{4\pi\epsilon_0|\vec{r}-\vec{r'}|}
\end{eqnarray}
and the \textit{potential energy of exchange and
  correlation}\index{exchange and correlation!potential energy}
\begin{eqnarray}
U_{xc}\stackrel{\text{def}}{=}
\int d^3r\; n^{(1)}(\vec{r})
\frac{1}{2}\int d^3r'\; \frac{e^2\;h(\vec{r},\vec{r'})}
{4\pi\epsilon_0|\vec{r}-\vec{r'}|}\;.
\label{eq:uxc}
\end{eqnarray}
Keep in mind that $U_{xc}$ is \textit{not} the exchange correlation
energy. The difference is a kinetic energy correction that will be
discussed later in Eq.~\ref{eq:exc}.

The hole function has a physical meaning: An electron sees the total
density minus the electrons accounted for by the hole. Thus each
electron does not only experience the electrostatic potential of the
total electron density $n^{(1)}(\vec{r})$, but also the attractive
potential of its own exchange correlation hole $h(\vec{r}_0,\vec{r})$.

\begin{figure}[h]
 \centering
 \includegraphics[width=0.4\textwidth]{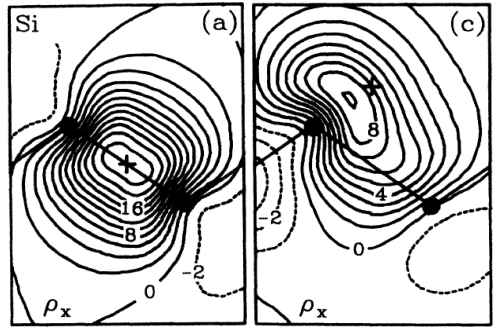}
 \caption{Exchange hole in silicon. The cross indicates the position of
   the observer electron. The black spheres and the lines indicate the
   atomic positions and bonds in the (110) plane. Reprinted figure
   with permission from Mark S. Hybertsen and Steven G. Louie,
   Physical Review B 34, 5390 (1986). Copyright 1986 by the American
   Physical Society}
\end{figure}

A few facts for this hole density are apparent:
\begin{enumerate}
\item Because each electron of a N-electron system sees $N-1$ other
  electrons, the hole function integrates to exactly minus one electron
\begin{eqnarray}
\int d^3r\;h(\vec{r}_0,\vec{r})=-1
\end{eqnarray}
irrespective of the position $\vec{r}_0$ of the observing electron.
\item The density of the remaining $N-1$ electrons can not be larger
  than the total electron density. This implies
\begin{eqnarray}
h(\vec{r}_0,\vec{r})\ge-n^{(1)}(\vec{r}) \;.
\end{eqnarray}
\item Due to the Pauli principle, no other electron with the same spin
  as the observer electron can be at the position $\vec{r}_0$. Thus
  the on-top hole $h(\vec{r}_0,\vec{r}_0)$ obeys the
  limits \cite{burke98_jcp109_3760}
\begin{eqnarray}
-\frac{1}{2}n^{(1)}(\vec{r}_0)
\ge h(\vec{r}_0,\vec{r}_0)
\ge -n^{(1)}(\vec{r}_0)\;.
\end{eqnarray}
\item Assuming locality, the hole function vanishes at large distances
  from the observer electron at $\vec{r}_0$, i.e.
\begin{eqnarray}
h(\vec{r}_0,\vec{r})\rightarrow0
\qquad\text{for}\qquad|\vec{r}-\vec{r}_0|\rightarrow\infty\;.
\end{eqnarray}
With locality I mean that the density does not depend on the position
or the presence of an observer electron, if the latter is very far
away.
\end{enumerate}

\subsection*{A selfmade functional}
\begin{figure}[h]
\begin{minipage}[t]{0.4\linewidth}
\begin{center}
 \includegraphics[width=\textwidth]{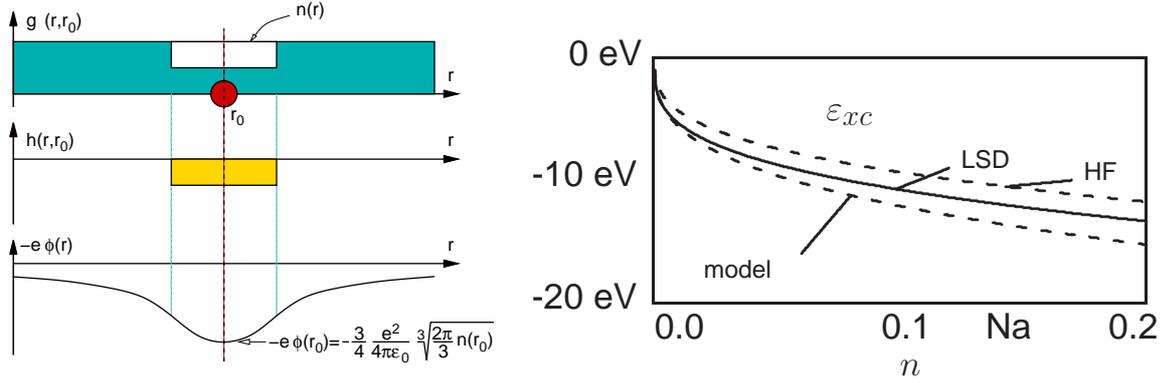}
\end{center}
\end{minipage}
\begin{minipage}[t]{0.4\linewidth}
\begin{center}
\input{fig2a.tex}
\end{center}
\end{minipage}
\caption{\label{fig:modelxc}Left: Scheme to demonstrate the
  construction of the exchange correlation energy from a simple
  model. Right: exchange correlation energy per electron
  $\epsilon_{xc}$ as function of electron density from our model,
  Hartree-Fock approximation and the exact result. The symbol ``Na''
  indicates the density of Sodium.}
\end{figure}

It is fairly simple to make our own density functional\footnote{A
  functional $F[y]$ maps a function $y(x)$ to a number $F$. It is a
  generalization of the function $F(\vec{y})$ of a vector $\vec{y}$,
  where the vector index of $\vec{y}$ is turned into a continuous
  argument $x$.}\index{functional}: For a given density, we choose a
simple shape for the hole function, such as a spherical box. Then we
scale the value and the radius such that the hole function integrates
to $-1$, and that its value is opposite equal to the spin density at
its center. The electrostatic potential of this hole density at its
center is the exchange-correlation energy for the observer
electron. Our model has an exchange correlation energy\footnote{For
  this model we do not distinguish between the energy of exchange and
  correlation and its potential energy contribution} of
\begin{eqnarray}
U_{xc}[n^{(1)}]\approx-\frac{1}{2}\int d^3r\; n^{(1)}(\vec{r})
\biggl(
\frac{3}{4}\frac{e^2}{4\pi\epsilon_0}
\sqrt[3]{\frac{2\pi}{3}}\biggl(n^{(1)}(\vec{r})\biggr)
^{\frac{1}{3}}\biggr)
\sim \int d^3r\;\biggl(n^{(1)}(\vec{r})\biggr)^{\frac{4}{3}}\;.
\end{eqnarray}

The derivation is an elementary exercise and is given in the
appendix. The resulting energy per electron $\epsilon_{xc}$ is given
on the right-hand side of Fig.~\ref{fig:modelxc} indicated as
``model'' and compared with the exact result indicated as ``LSD'' and
the Hartree-Fock result indicated as ``HF'' for a homogeneous electron
gas.

The agreement with the correct result, which is surprisingly good for
such a crude model, provides an idea of how robust the
density-functional theory is with respect to approximations.  While
this model has been stripped to the bones, it demonstrates the way
physical insight enters the construction of density
functionals. Modern density functionals are far more sophisticated and
exploit much more information \cite{perdew96_ijqc57_309}, but the basic
method of construction is similar.

\subsection*{Kinetic energy}
While the expression for the kinetic energy in
Eq.~\ref{eq:etotwithnatorbandn2} seems familiar, there is a catch to
it. In order to know the natural orbitals and the occupations we need
access to the many-particle wave function or at least to its reduced
density matrix.

A good approximation for the kinetic energy of the interacting
electrons is the kinetic energy functional\index{kinetic energy
  functional} $T_s[n^{(1)}]$ of the ground state of non-interacting
electrons with the same density as the true system. It is defined by
\begin{eqnarray}
  T_s[n^{(1)}]
  &=&\hspace{-0.3cm}\min_{\{f_n\in[0,1],|\psi_n\rangle\}}
  \biggl\{
  \sum_n f_n\int d^4x\; 
  \psi^*_n(\vec{x})\frac{-\hbar^2\vec{\nabla^2}}{2m}\psi_n(\vec{x})
  \nonumber\\
&&\hspace{2cm}  +\int d^3r\;
  v_{eff}(\vec{r})\biggl(\biggl[\sum_nf_n
  \sum_\sigma\psi_n^*(\vec{x})\psi_n(\vec{x})\biggr]
  -n^{(1)}(\vec{r})\biggr)
\nonumber\\
&&\hspace{2cm} -\sum_{n,m}\Lambda_{m,n}
\biggl(\langle\psi_n|\psi_m\rangle-\delta_{n,m}\biggr)
  \biggr\}\;.
\end{eqnarray}
Note that $f_n\neq \bar{f}_n$ and that the so-called Kohn-Sham
orbitals\index{Kohn-Sham orbitals} $\psi_n(\vec{x})$
differ\footnote{To be precise, Kohn-Sham orbitals are the natural
  orbitals for non-interacting electrons of a given density. They are
  however different from the natural orbitals of interacting electrons
  at the same density.} from the natural orbitals
$\varphi_n(\vec{x})$.  Natural orbitals and Kohn-Sham wave functions
are fairly similar, while the occupations $f_n$ of Kohn-Sham orbitals
differ considerably from those $\bar{f}_m$ of the natural orbitals.
The effective potential $v_{eff}(\vec{r})$ is the Lagrange multiplier
for the density constraint. $\Lambda_{m,n}$ is the Lagrange multiplier
for the orthonormality. Diagonalization of $\bm\Lambda$ yields a
diagonal matrix with the one-particle energies on the diagonal.

This kinetic energy $T_s[n^{(1)}]$ is a unique functional of the density,
which is the first sign that we are approaching a
density-functional theory. Also it is the introduction of this kinetic
energy, where we made for the first time a reference to a ground
state. Density functional theory as described here is inherently a
ground-state theory.

Why does the true kinetic energy of the interacting system differ from
that of the non-interacting energy? Consider the hole function of a
non-interacting electron gas. When inserted into Eq.~\ref{eq:uxc} for
$U_{xc}$ the potential energy of exchange and correlation, we obtain a
contribution to the total energy that is called exchange
energy\index{exchange energy}. The interaction leads to a second
energy contribution that is called correlation
energy\index{correlation energy}. Namely, when the interaction is
switched on, the wave function is deformed in such a way that the
Coulomb repulsion between the electrons is reduced.  This makes the
hole function more compact. However, there is a price to pay when the
wave functions adjust to reduce the Coulomb repulsion between the
electrons, namely an increase of the kinetic energy: Pushing electrons
away from the neighborhood of the reference electrons requires to
perform work against the kinetic pressure of the electron gas, which
raises the kinetic energy.  Thus, the system has to find a
compromise between minimizing the electrostatic repulsion of the
electrons and increasing its kinetic energy. As a result, the
correlation energy has a potential-energy contribution and a
kinetic-energy contribution.

This tradeoff can be observed in Fig.~\ref{fig:modelxc}. The correct
exchange correlation energy is close to our model at low densities,
while it becomes closer to the Hartree-Fock result at high
densities. This is consistent with the fact that the electron gas can
easily be deformed at low densities, while the deformation becomes
increasingly costly at high densities due to the larger pressure of
the electron gas.

The difference between $T_s$ and the true kinetic energy is combined
with the potential energy of exchange and correlation $U_{xc}$ from
Eq.~\ref{eq:uxc} into the exchange correlation energy\index{exchange
  and correlation energy} $E_{xc}$, i.e.
\begin{eqnarray}
E_{xc}=U_{xc}+\sum_n \bar{f}_n \int d^4x\; 
\varphi_n^*(\vec{x})\frac{-\hbar^2}{2m}\vec{\nabla}^2\varphi_n(\vec{x})
-T_s[n^{(1)}]\;.
\label{eq:exc}
\end{eqnarray}
Note, that the $\phi_n(\vec{x})$ and the $\bar{f}_n$ are natural
orbitals and occupations of the interacting electron gas, and that
they differ from the Kohn-Sham orbitals $\psi_n(\vec{x})$ and
occupations $f_n$.

\subsection*{Total energy}
The total energy obtains the form
\begin{eqnarray}
  E&=&\min_{|\Phi\rangle,\{|\psi_n\rangle,f_n\in[0,1]\}}
  \biggl\{\sum_n f_n \int d^4x\; 
  \psi_n^*(\vec{x})\frac{-\hbar^2}{2m}\vec{\nabla}^2\psi_n(\vec{x})
  \nonumber\\
  &+&\int d^3r\;
  v_{eff}(\vec{r})\biggl(\biggl[\sum_n 
f_n \sum_\sigma\psi_n^*(\vec{x})\psi_n(\vec{x})\biggr]
  -n(\vec{r})\biggr)
  +\int d^3r\; v_{ext}(\vec{r})n^{(1)}(\vec{r})
  \nonumber\\
  &+&\frac{1}{2}\int d^3r\int d^3r'\;
  \frac{e^2n^{(1)}(\vec{r})n^{(1)}(\vec{r'})}
  {4\pi\epsilon_0|\vec{r}-\vec{r'}|}+E_{xc}
  -\sum_{n,m}\Lambda_{m,n}\biggl(\langle\psi_n|\psi_m\rangle-\delta_{n,m}\biggr)
  \biggr\}\;.
\label{eq:etot2}
\end{eqnarray}
In order to evaluate the total energy with Eq.~\ref{eq:etot2}, we
still have to start from the many-particle wave function
$|\Phi\rangle$. Only the many-particle wave function allows us to
evaluate the one-particle density $n^{(1)}(\vec{r})$ and the exchange
correlation energy $E_{xc}$.  Kohn-Sham orbitals $|\psi_n\rangle$ and
occupations $f_n$ are obtained by an independent minimization
for each density.

If, however, we were able to express the exchange-correlation energy
$E_{xc}$ as a functional of the density alone, there would be no need
for the many-particle wave function at all and the terrors of the
exponential wall would be banned. We could minimize Eq.~\ref{eq:etot2}
with respect to the density, Kohn-Sham orbitals and their occupations.

Let us, for the time being, simply assume that $E_{xc}[n^{(1)}]$ is a
functional of the electron density and explore the consequences of
this assumption.  Later, I will show that this assumption is actually
valid.

The minimization in Eq.~\ref{eq:etot2} with respect to the
one-particle wave functions yields the Kohn-Sham equations
\begin{eqnarray}
\left[\frac{-\hbar^2}{2m_e}\vec{\nabla}^2
+v_{eff}(\vec{r})-\epsilon_n\right]\psi_n(\vec{x})=0
\qquad\text{with}\qquad 
\int d^4x\;\psi_m(\vec{x})\psi_n(\vec{x})=\delta_{m,n}\;.
\label{eq:kohnshameq}
\end{eqnarray}
The Kohn-Sham energies\index{Kohn-Sham energy} $\epsilon_n$ are the
diagonal elements of the Lagrange multiplier ${\bf\Lambda}$, when the
latter is forced to be diagonal.

The requirement that the derivative of the total energy
Eq.~\ref{eq:etot2} with respect to the density vanishes, yields an
expression for the effective potential
\begin{eqnarray}
v_{eff}(\vec{r})=v_{ext}(\vec{r})
+\int d^3r'\;
\frac{e^2n^{(1)}(\vec{r'})}
{4\pi\epsilon_0|\vec{r}-\vec{r'}|}
+\frac{\delta E_{xc}[n^{(1)}]}{\delta n^{(1)}(\vec{r})}\;.
\label{eq:veff}
\end{eqnarray}
Both equations, together with the density constraint
\begin{eqnarray}
n^{(1)}(\vec{r})=\sum_n f_n \sum_\sigma\psi_n^*(\vec{x})\psi_n(\vec{x})\;,
\label{eq:densityfromkohnsham}
\end{eqnarray}
form a set of coupled equations, that determine the electron density
and the total energy. This set of coupled equations,
Eqs.~\ref{eq:kohnshameq}, \ref{eq:veff}, and
\ref{eq:densityfromkohnsham}, is what is solved in the so-called
self-consistency loop.
Once the set of self-consistent equations has been solved, we obtain
the electron density and we can evaluate the total energy. 

\begin{figure}{h}
\begin{center}
\includegraphics[width=0.7\linewidth]{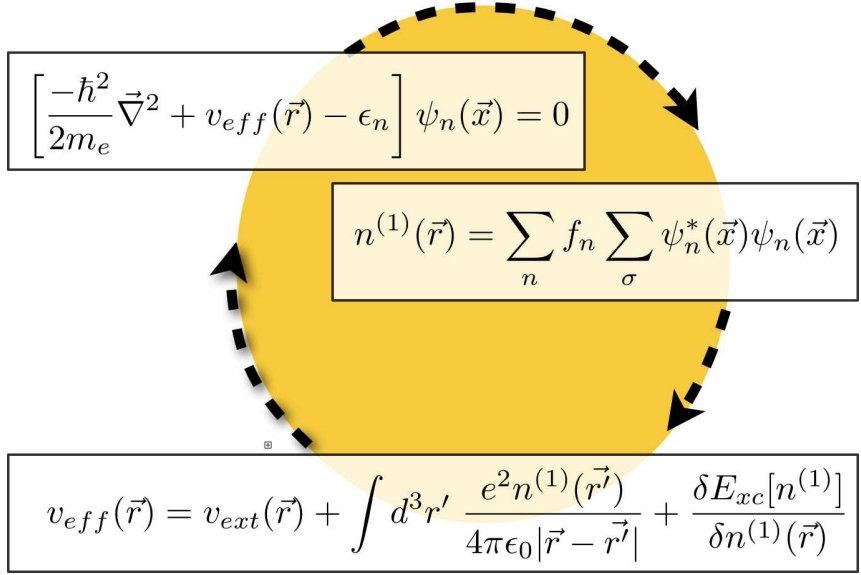}
\end{center}
\caption{Self-consistency cycle}
\end{figure}

In practice, one often makes the assumption that the non-interacting
electrons in the effective potential closely resemble the true
interacting electrons, and extracts a wealth of other physical
properties from the Kohn-Sham wave functions $|\psi_n\rangle$ and the
Kohn-Sham energies $\epsilon_n$. However, there is little theoretical
backing for this approach and, if it fails, one should not blame
density functional theory!

\subsection*{Is there a density functional?}
The argument leading to the self-consistent equations,
Eqs.~\ref{eq:kohnshameq}, \ref{eq:veff}, and
\ref{eq:densityfromkohnsham}, relied entirely on the hope that
exchange correlation functional can be expressed as a functional of
the electron density. In fact, this can easily be shown, if we
restrict us to ground state densities. The proof goes back to the
seminal paper by Levy \cite{levy79_pnas76_6062,lieb83_ijqc24_243}.

Imagine that one could construct all fermionic many-particle wave
functions. For each of these wave functions, we can determine in a
unique way the electron density 
\begin{eqnarray}
n^{(1)}(\vec{r})=N\sum_{\sigma}\int d^3x_2\ldots\int d^3x_N\;
|\Psi(\vec{x},\vec{x}_2,\ldots\vec{x}_N)|^2\;.
\end{eqnarray}
Having the electron densities, we sort the wave functions according to
their density. For each density, I get a mug $M[n^{(1)}]$ that holds
all wave functions with that density, that is written on the label of
the mug. 

\begin{figure}{h}
\begin{center}
\includegraphics[width=0.7\linewidth]{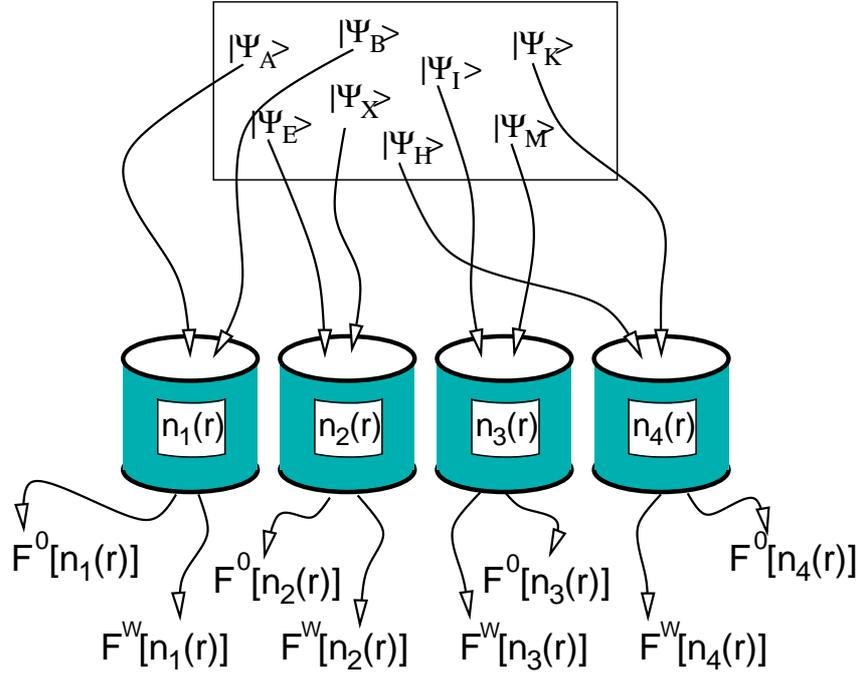}
\end{center}
\caption{Illustration for Levy's proof that there exists a density
  functional}
\end{figure}

Now we turn to each mug $M[n^{(1)}]$ in sequence and determine for
each the wave function with the lowest energy. Because the external
potential energy is the same for all wave functions with the same
density, we need to consider only the kinetic energy operator
$\hat{T}$ and the operator $\hat{W}$ of the electron-electron
interaction, and we do not need to consider the external potential.
\begin{eqnarray}
F^{\hat{W}}[n^{(1)}]=\min_{|\Psi\rangle\in M[n^{(1)}]}
\langle\Psi|\hat{T}+\hat{W}|\Psi\rangle
\end{eqnarray}
$F^{\hat{W}}[n^{(1)}]$ is the universal density functional. It is
universal in the sense that it is an intrinsic property of the
electron gas and absolutely independent of the external potential.

Next, we repeat the same construction as that for a universal density
functional, but now we leave out the interaction $\hat{W}$ and 
consider only the kinetic energy $\hat{T}$.
\begin{eqnarray}
F^{0}[n^{(1)}]=\min_{|\Psi\rangle\in M[n^{(1)}]}
\langle\Psi|\hat{T}|\Psi\rangle
\end{eqnarray}
The resulting functional $F^{0}[n^{(1)}]$ is nothing but the kinetic
energy of non-interacting electrons $T_s[n^{(1)}]$.

Now we can write down the total energy as functional of the density
\begin{eqnarray}
  E[n^{(1)}]&=& F^{\hat{W}}[n^{(1)}]
  +\int d^3r\; v_{ext}(\vec{r})n^{(1)}(\vec{r})
\label{eq:etotwithfunctional}
\end{eqnarray}

When we compare Eq.~\ref{eq:etotwithfunctional} with Eq.~\ref{eq:etot2}, 
we obtain  an expression for the exchange
correlation energy.
\begin{eqnarray}
  E_{xc}[n^{(1)}]&=& F^{\hat{W}}[n^{(1)}(\vec{r})]-F^{0}[n^{(1)}(\vec{r})]
-\frac{1}{2}\int d^3r\int d^3r'\;
\frac{e^2n^{(1)}(\vec{r})n^{(1)}(\vec{r'})}
{4\pi\epsilon_0|\vec{r}-\vec{r'}|}
\label{eq:defexc}
\end{eqnarray}
This completes the proof that the exchange correlation energy is a
functional of the electron density. The latter was the assumption for
the derivation of the set of self-consistent equations,
Eqs.~\ref{eq:kohnshameq}, \ref{eq:veff}, and \ref{eq:densityfromkohnsham}
for the Kohn-Sham wave functions $\psi_n(\vec{x})$.

With this, I finish the description of the theoretical basis of
density-functional theory. We have seen that the total energy can
rigorously be expressed as a functional of the density or, in
practice, as a functional of a set of one-particle wave functions, the
Kohn-Sham wave functions and their occupations. Density functional
theory per se is not an approximation and, in contrast to common
belief, it is not a mean-field approximation. Nevertheless, we need to
introduce approximations to make density functional theory work. This
is because the exchange correlation energy $E_{xc}[n^{(1)}]$ is not
completely known. These approximations will be discussed in the next
section.

\section{Jacob's ladder of density functionals}
The development of density functionals is driven by mathematical
analysis of the exact exchange correlation hole
\cite{perdew96_ijqc57_309,perdew05_jcp123_62201}, physical insight and
numerical benchmark calculations on real systems. The functionals
evolved in steps from one functional form to another, with several
parameterizations at each level. Perdew pictured this development by
Jacob's ladder leading up to heaven
\cite{perdew01_aipcp577_1,perdew05_jcp123_62201}. In his analogy the
different rungs of the ladder represent the different levels of
density functionals leading to the unreachable, ultimately correct
functional.

\subsection*{LDA, the big surprise}
The first density functionals used in practice were based on the
local-density approximation (LDA)\index{local density
  approximation}\index{LDA}. The hole function for an electron at
position $\vec{r}$ has been approximated by the one of a homogeneous
electron gas with the same density as $n^{(1)}(\vec{r})$. The exchange
correlation energy for the homogeneous electron gas has been obtained
by quantum Monte Carlo calculations \cite{ceperley80_prl45_566} and
analytic calculations \cite{ma68_pr165_18}.  The local density
approximation has been generalized early to local spin-density
approximation (LSD)\index{local spin density approximation}\index{LSD}
\cite{vonbarth72_jpcss5_1629}.

Truly surprising was how well the theory worked for real systems.
Atomic distances could be determined within a few percent of the bond
length and energy differences in solids were surprisingly good.

This was unexpected, because the density in real materials is far from
homogeneous. Gunnarsson and Lundquist \cite{gunnarsson76_prb13_4274}
explained this finding with sumrules, that are obeyed by the local
density approximation: Firstly, the exchange correlation energy
depends only on the spherical average of the exchange correlation
hole. Of the radial hole density only the first moment contributes,
while the second moment is fixed by the sum-rule that the electron
density of the hole integrates to $-1$. Thus we can use
\begin{eqnarray}
\int d^3r\; \frac{e^2h(\vec{r}_0,\vec{r})}{4\pi\epsilon_0|\vec{r}-\vec{r}_0|}
=-\frac{e^2}{4\pi\epsilon_0}
\frac{\int_0^\infty dr\; r 
\bigl\langle h(\vec{r}_0,\vec{r'})\bigr\rangle_{|\vec{r'}-\vec{r}_0|=r}}
{\int_0^\infty dr\; r^2 
\bigl\langle h(\vec{r}_0,\vec{r'})\bigr\rangle_{|\vec{r'}-\vec{r}_0|=r}}
\end{eqnarray}
where the angular brackets imply the angular average of
$\vec{r'}-\vec{r}_0$.  This dependence on the hole density is rather
insensitive to small changes of the hole density. Even for an atom,
the \textit{spherically averaged} exchange hole closely resembles that
of the homogeneous electron gas \cite{jones89_rmp61_689}.

The main deficiency of the LDA was the strong overbinding with bond
energies in error by about one electron volt. On the one hand, this
rendered LDA useless for most applications in chemistry. On the other
hand, the problem was hardly visible in solid state physics where
bonds are rarely broken, but rearranged so that errors cancelled.

\subsection*{GGA, entering chemistry}
Being concerned about the large density variations in real materials,
one tried to include the first terms of a Taylor expansion in the
density gradients.  These attempts failed miserably. The culprit has
been a violation of the basic sum rules as pointed out by Perdew
\cite{perdew85_prl55_1665}. The cure was a cutoff for the gradient
contributions at high gradients, which lead to the class of
generalized gradient approximations (GGA)\index{generalized gradient
  approximation}\index{GGA} \cite{langreth83_prb28_1809}.

Becke \cite{becke88_pra38_3098} provides an intuitive description for
the workings of GGA's, which I will sketch here in a simplified
manner: Becke uses an ansatz $E_{xc}=\int d^3r\;
A(n(\vec{r}))F(x(\vec{r}))$ for the exchange-correlation energy where
$n(\vec{r})$ is the local density and
$x=|\vec{\nabla}n|/n^{\frac{4}{3}}$ is a dimensionless reduced
gradient. Do not confuse this symbol with the combined
position-and-spin coordinate $\vec{x}$. The function $A$ is simply the
LDA expression and $F(x)$ is the so-called enhancement factor.  The
large-gradient limit of $F(x)$ is obtained from a simple physical
argument:

\begin{figure}[h]
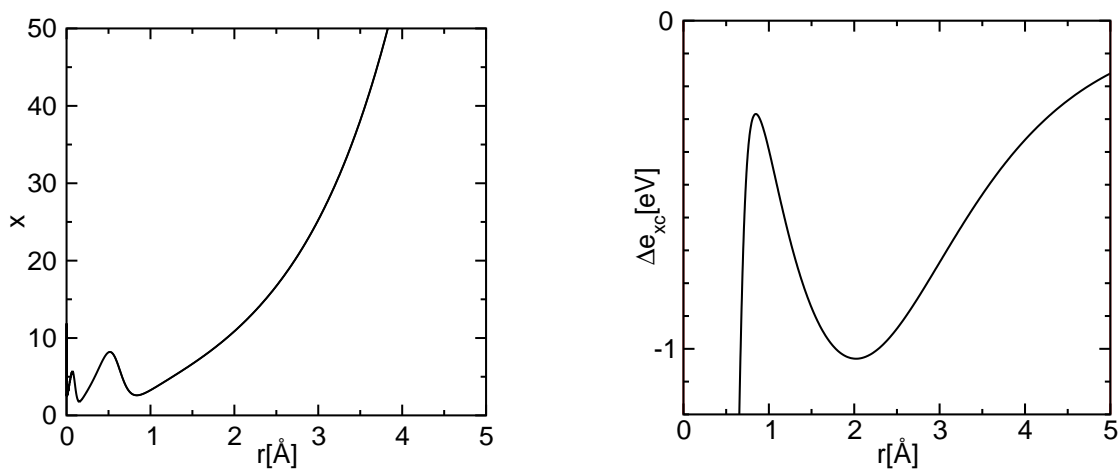

\begin{center}
\includegraphics[width=0.4\linewidth,clip=true]{fig5a.eps}
\hspace{0.1\linewidth}
\includegraphics[width=0.4\linewidth,clip=true]{fig5b.eps}
\end{center}
\caption{\label{fig:gradientanalysis}Left figure: reduced density
  gradient $x=|\vec{\nabla} n|/n^{\frac{4}{3}}$ of a silicon atom as function
    of distance from the nucleus demonstrating that the largest
    reduced gradients occur in the exponential tails.  Right figure:
    additional contribution from the gradient correction (PBE versus
    PW91 LDA) of the exchange correlation energy per electron. The
    figure demonstrates that the gradient correction stabilizes the
    tails of the wave function. The covalent radius of silicon is at
    1.11~\AA.}
\end{figure}

Somewhat surprisingly, the reduced gradient is largest not near the
nucleus but in the exponentially decaying charge-density tails as
shown in Fig.~\ref{fig:gradientanalysis}.  For an electron that is far
from an atom, the hole is on the atom, because a hole can only be
dug where electrons are. Thus the Coulomb interaction energy of the
electron with its hole is $-\frac{e^2}{4\pi\epsilon_0 r}$,
where $r$ is the distance of the reference electron from the atom. As
shown in appendix \ref{app:fofx}, the enhancement factor can now be
obtained by enforcing this behavior for exponentially decaying
densities.

As a result, the exchange and correlation energy per electron in the
tail region of the electron density falls of with the inverse distance
in GGA, while it has a much faster, exponential decay in the
LDA. Thus, the the tail region is stabilized by GGA. This contribution
acts like a negative ``surface energy''. 

When a bond between two atoms is broken, the surface is increased. In
GGA this bond-breaking process is more favorable than in LDA, and,
hence, the bond is weakened. Thus the GGA cures the overbinding
error of the LDA.

These gradient corrections greatly improved the bond energies and made
density functional theory useful also for chemists.  The most widely
distributed GGA functional is the Perdew-Burke-Ernzerhof (PBE)
functional \cite{perdew96_prl77_3865}.

\subsection*{Meta GGA's}
The next level of density functionals are the so-called meta
GGA's\index{meta GGA} \cite{pryonov95_ijqc56_61, voorhis98_jcp109_400,
  becke98_jcp109_2092} that include not only the gradient of the
density, but also the second derivatives of the density. These
functionals can be reformulated so that the additional parameter is
the kinetic energy density instead of the second density derivatives.
Perdew recommends his TPSS functional \cite{tao03_prl91_146401}.

\subsection*{Hybrid functionals}
Another generation of functionals are hybrid functionals\index{hybrid
  functional} \cite{becke93_jcp98_1372,becke93_jcp98_5648}, which
replace some of the exchange energy by the exact exchange
\begin{eqnarray}
E_X^{HF}=-\frac{1}{2}\sum_{m,n}\bar{f}_m\bar{f}_n\int d^4x\int d^4x'
\frac{e^2\psi^*_m(\vec{x})\psi_n(\vec{x})
\psi^*_n(\vec{x'})\psi_m(\vec{x'})}{4\pi\epsilon_0|\vec{r}-\vec{r'}|}
\label{eq:ex}
\end{eqnarray}
where $\bar{f}_n$ and the $\psi_n(\vec{x})$ are the Kohn-Sham
occupations and wave functions, respectively.

The motivation for this approach goes back to the adiabatic connection
formula\index{adiabatic connection}
\cite{harris74_jpf4_1170,langreth75_ssc17_1425,
  gunnarsson76_prb13_4274}
\begin{eqnarray}
  E_{xc}[n(\vec{r})]
=\int_0^1 d\lambda\;U_{xc}^{\lambda\hat{W}}[n(\vec{r})]
=
\int d^3r\; n(\vec{r})\int_0^1 
  d\lambda 
  \frac{1}{2}\int d^3r'
\frac{h_\lambda(\vec{r},\vec{r'})}{4\pi\epsilon|\vec{r}-\vec{r'}|}
\label{eq:excadiabatic}
\end{eqnarray}
which expresses the exchange correlation energy as an integral of the
potential energy of exchange and correlation over the interaction
strength. Here the interaction in the Hamiltonian is scaled by a
factor $\lambda$, leading to a $\lambda$-dependent universal
functional $F^{\lambda\hat{W}}[n^{(1)}]$. The interaction energy can
be expressed by
\begin{eqnarray}
F^{\hat{W}}[n]
&=&F^{0}[n]
+\int_0^1d\lambda\;\frac{d}{d\lambda}F^{\lambda\hat{W}}[n]
\nonumber\\
&=&T_s[n]
+\frac{1}{2}\int d^3r\int d^3r'\;
\frac{e^2n(\vec{r})n(\vec{r'})}{4\pi\epsilon_0|\vec{r}-\vec{r'}|}
+\int_0^1 d\lambda\; U_{xc}^{\lambda\hat{W}}[n]
\end{eqnarray}
which leads via Eq.~\ref{eq:defexc} to Eq.~\ref{eq:excadiabatic}.
Using perturbation theory, the derivative of $F^{\lambda\hat{W}}[n]$
simplifies to the expectation
$\langle\Psi(\lambda)|\hat{W}|\Psi(\lambda)\rangle$ value of the
interaction, which is the potential energy of exchange and correlation
evaluated for a many-particle wave function obtained for the specified
given interaction strength.

The underlying idea of the hybrid functionals is to interpolate the
integrand between the end points. In the non-interacting limit,
i.e. for $\lambda=0$ the integrand $U_{xc}^{\lambda\hat{W}}$ is
exactly given by the exact exchange energy of Eq.~\ref{eq:ex}. For the
full interaction, on the other hand, the LDA or GGA functionals are
considered correctly. Thus a linear interpolation would yield
\begin{eqnarray}
E_{xc}=\frac{1}{2}\biggl(U_{xc}^{0}+U_{xc}^{\hat{W}}\biggr)
=\frac{1}{2}\biggl(E_{X}^{HF}+U_{xc}^{\hat{W}}\biggr)
=E_{xc}^{GGA}+\frac{1}{2}\biggl(E_{X}^{HF}-E_X^{GGA}\biggr)\,.
\end{eqnarray}
Depending on whether the $\lambda$-dependence is a straight line or
whether it is convex, the weight factor may be equal or smaller than
$\frac{1}{2}$. Perdew \cite{perdew96_jcp105_9982} has given arguments
that a factor $\frac{1}{4}$ would actually be better than a factor
$\frac{1}{2}$.

Hybrid functionals perform substantially better than GGA functionals
regarding binding energies, band gaps and reaction energies. However,
they are flawed for the description of solids. The reason is that the
exact exchange hole in a solid is very extended. These long-range
tails are screened away quickly when the interaction is turned on,
because they are cancelled by the correlation. Effectively, we should
use a smaller mixing factor for the long range part of the exchange
hole. This can be taken into account, by cutting off the long-range
part of the interaction for the calculation of the Hartree-Fock
exchange \cite{heyd03_jcp118_8207}.  This approach improves the
results for band gaps while reducing the computational effort
\cite{marsman08_jpc20_64201}.

The effective cancellation of the long-ranged contribution of exchange
with a similar contribution from correlation, which is also considered
properly already in the LDA, is one of the explanation for the
superiority of the LDA over the Hartree-Fock approximation.

The most widely used hybrid functional is the B3LYP functional
\cite{stephens94_jpc98_11623}, which is, however, obtained from a
parameter fit to a database of simple molecules. The functional
PBE0 \cite{ernzerhof99_jcp110_5029, adamo99_jcp110_6158} is born out of
the famous PBE GGA functional and is a widely distributed
parameter-free functional.

\subsection*{LDA+U and local hybrid functionals}
Starting from a completely different context, Anisimov
et. al. \cite{anisimov91_prb44_943} introduced the so-called LDA+U
method, which, as described below, has some similarities to the
hybrid functionals above.

The main goal was to arrive at a proper description of transition
metal oxides, which tend to be Mott insulators, while GGA calculations
predict them often to be metals. The remedy was to add a correlation
term\footnote{The expression given here looks unusually simple. This
  is due to the notation of spin orbitals, which takes care of the
  spin indices.}  \cite{liechtenstein95_prb52_5467} borrowed from the
Hubbard model and to correct the resulting double counting of the
interactions by $E_{dc}$.
\begin{eqnarray}
  E&=&E^{GGA}+\frac{1}{2}\sum_R \sum_{\alpha,\beta,\gamma,\delta\in\mathcal{C}_R}
  U_{\alpha,\beta,\gamma,\delta} 
  \biggl(\rho_{\gamma,\alpha}\rho_{\delta,\beta}
  -\rho_{\delta,\alpha}\rho_{\gamma,\beta}\biggr)-E_{dc}
  \\
  U_{\alpha,\beta,\gamma,\delta}
  &=&\int d^4x\int d^4x'
  \frac{e^2
    \chi^*_\alpha(\vec{x})\chi^*_\beta(\vec{x'})
    \chi_\gamma(\vec{x})\chi_\delta(\vec{x'})}
  {4\pi\epsilon_0|\vec{r}-\vec{r'}|}
  \\
  \rho_{\alpha,\beta}&=&\langle\pi_\alpha|\psi_n \rangle 
  f_n\langle\psi_n|\pi_\beta\rangle
\end{eqnarray}
where $|\chi_\alpha\rangle$ are atomic tight-binding orbitals and
$|\pi_\alpha\rangle$ are their projector functions.\footnote{Projector
  functions obey the biorthogonality condition
  $\langle\chi_\alpha|\pi_\beta\rangle=\delta_{\alpha,\beta}$. Within
  the sub-Hilbert space of the tight-binding orbitals, i.e. for wave
  functions of the form $|\psi\rangle=\sum_\alpha|\chi_\alpha\rangle
  c_\alpha$, the projector functions decompose the wave function into
  tight binding orbitals, i.e. $|\psi\rangle=\sum_\alpha
  |\chi_\alpha\rangle\langle\pi_\alpha|\psi\rangle$. A similar
  projection is used extensively in the projector augmented-wave
  method described later.} The additional energy is a Hartree-Fock
exchange energy, that only considers the exchange for specified sets
of local orbitals. The exchange term does only consider a subset of
orbitals $\mathcal{C}_R$ for each atom $R$ and it ignores the
contribution involving orbitals centered on different atoms.

Novak et al. \cite{novak06_pss243_563} made the connection to the
hybrid functionals explicit and restricted the exact exchange
contribution of a hybrid functional to only a shell of orbitals. While
in the LDA+U method the bare Coulomb matrix elements are reduced by a
screening factor, in the hybrid functionals it is the mixing factor
that effectively plays the same role. Both, LDA+U and the local hybrid
method, have in common that they radically remove the contribution of
off-site matrix elements of the interaction. Tran et
al. \cite{tran06_prb74_155108} applied this method to transition metal
oxides and found results that are similar to those of the full
implementation of hybrid functionals.

\subsection*{Van der Waals interactions}
One of the major difficulties for density functionals is the
description of van der Waals forces, because it is due to the quantum
mechanical synchronization of charge fluctuations on distinct
molecules.  I refer the reader to the work made in the group of
Lundqvist \cite{dion04_prl92_246401, thonhauser07_prb76_125112,
  lee10_prb82_81101}.

\section{Benchmarks,  successes and failures}
The development of density functionals has profited enormously on
careful benchmark studies. The precondition is a data set of test
cases for which reliable and accurate experimental data exist. The
most famous data sets are the G1 and G2 databases
\cite{pople89_jcp90_5622, curtiss90_jcp93_2537, curtiss97_jcp106_1063,
  curtiss98_jcp109_42} that have been set up to benchmark
quantum-chemistry codes.  Becke \cite{becke92_jcp96_2155,
  becke92_jcp97_9173, becke93_jcp98_5648, becke96_jcp104_1040,
  becke97_jcp107_8554} set a trend by using these large sets of test
cases for systematic studies of density functionals.  In order to
separate out the accuracy of the density functionals, it is vital to
perform these calculations on extremely accurate numerical
methods. Becke used basis set free calculations, that were limited to
small molecules while being extremely accurate. Paier et.
al. \cite{paier05_jcp122_234102,paier06_jcp124_154709,
  paier06_jcp125_249901, marsman08_jpc20_64201} have later performed
careful comparisons of two methods, Gaussian and the projector
augmented-wave method, to single out the error of the electronic
structure method.

Overall, the available density functionals predict molecular
structures very well. Bond distances agree with the experiment often
within one percent. Bond angles come out within a few degrees. 

The quality of total energies depends strongly on the level of
functionals used.  On the LDA level bonds are overestimated in the
1~eV range, on the GGA level these errors are reduced to a about
0.3~eV, and hybrid functionals reduce the error by another factor of
2. The ultimate goal is to reach chemical accuracy, which is about
0.05~eV. Such an accuracy allows to predict reaction rates at room
temperature within a factor of 10.

Band gaps are predicted too small with LDA and GGA. The
so-called band gap problem has been one of the major issues during the
development of density functionals. Hybrid functionals clearly improve
the situation. A problem is the description of materials with strong
electron correlations. For LDA and GGA many insulating transition
metal oxides are described as metals. This changes again for the
hybrid functionals, which turns them into antiferromagnetic
insulators, which is a dramatic improvement.

\section{Electronic structure methods}
In this second part of my lecture notes, I will address the problem
how to solve the Kohn-Sham equations and how to obtain the total
energy and other observables. It is convenient to use a slightly
different notation: Instead of treating the nuclei via an external
potential, we combine all electrostatic interactions into a single
double integral.

This brings the total energy into the form
\begin{eqnarray}
E\Bigl[\{\psi_n(\vec{r})\},\{\vec{R}_R\}\Bigr]&=&\sum_n f_n 
\langle\psi_n|\frac{\hat{\vec{p}}\,^2}{2m_e}|\psi_n\rangle
\nonumber\\
&&\hspace{-2cm}+\frac{1}{2}\int d^3r\int
d^3r'\;\frac{e^2\Bigl(n(\vec{r})+Z(\vec{r})\Bigr)
\left(n(\vec{r'})+Z(\vec{r'})\right)}
{4\pi\epsilon_0|\vec{r}-\vec{r'}|}
+E_{xc}[n]\;,
\label{eq:dfttotalenergy}
\end{eqnarray}
where $Z(\vec{r})=- \sum_R
\mathcal{Z}_R\delta(\vec{r}-\vec{R}_R)$ is the nuclear charge
density expressed in electron charges.  $Z_R$ is the
atomic number of a nucleus at position $\vec{R}_R$.

The electronic ground state is determined by minimizing the total
energy functional $E[\Psi_n]$ of Eq.~\ref{eq:dfttotalenergy} at a
fixed ionic geometry. The one-particle wave functions have to be
orthogonal. This constraint is implemented with the method of Lagrange
multipliers. We obtain the ground-state wave functions from the
extremum condition for
\begin{eqnarray}
Y\Bigl[\{|\psi_n\rangle\},\mat{\Lambda}\Bigr]=E\Bigl[\{|\psi_n\rangle\}\Bigr]
-\sum_{n,m}\Bigl[\langle\psi_n|\psi_m\rangle-\delta_{n,m}\Bigr]\Lambda_{m,n}
\label{eq:ewithconstraint}
\end{eqnarray}
with respect to the wavefunctions and the Lagrange multipliers
$\Lambda_{m,n}$.  The extremum condition for the wavefunctions has
the form
\begin{eqnarray}
\hat{H}|\psi_n\rangle f_n =\sum_m|\psi_m\rangle\Lambda_{m,n}\;,
\end{eqnarray}
where $\hat{H}=\frac{1}{2m_e}\hat{\vec{p}}\,^2+\hat{v}_{\mathrm{eff}}$ is the
effective one-particle Hamilton operator.

The corresponding effective potential depends itself on the electron
density via
\begin{eqnarray}
v_{eff}(\vec{r})&=&\int
d^3r'\;\frac{e^2\Bigl(n(\vec{r'})+
Z(\vec{r'})\Bigr)}{4\pi\epsilon_0|\vec{r}-\vec{r'}|}
+\mu_{xc}(\vec{r})\;,
\end{eqnarray}
where $\mu_{xc}(\vec{r})=\frac{\delta E_{xc}[n(\vec{r})]}{\delta
n(\vec{r})}$ is the functional derivative of the exchange and correlation
functional.

After a unitary transformation that diagonalizes the matrix of Lagrange
multipliers $\mat{\Lambda}$, we obtain the Kohn-Sham equations
\begin{eqnarray}
\hat{H}|\psi_n\rangle=|\psi_n\rangle\epsilon_n\;.
\label{eq:hpsigleichepsi}
\end{eqnarray}
The one-particle energies $\epsilon_n$ are the eigenvalues of the
matrix with the elements $\Lambda_{n,m}(f_n+f_m)/(2f_nf_m)$
\cite{bloechl94_prb50_17953}.

The one-electron Schr\"odinger equations, namely the Kohn-Sham
equations given in Eq.~\ref{eq:kohnshameq}, still pose substantial
numerical difficulties: (1) in the atomic region near the nucleus, the
kinetic energy of the electrons is large, resulting in rapid
oscillations of the wavefunction that require fine grids for an
accurate numerical representation.  On the other hand, the large
kinetic energy makes the Schr\"odinger equation stiff, so that a
change of the chemical environment has little effect on the shape of
the wavefunction.  Therefore, the wavefunction in the atomic region
can be represented well already by a small basis set.  (2) In the
bonding region between the atoms the situation is opposite.  The
kinetic energy is small and the wavefunction is smooth.  However, the
wavefunction is flexible and responds strongly to the
environment. This requires large and nearly complete basis sets.

Combining these different requirements is non-trivial and various
strategies have been developed.
\begin{itemize}
\item The atomic point of view has been most appealing to quantum
  chemists.  Basis functions are chosen that resemble atomic
  orbitals. This choice exploits that the wavefunction in the atomic
  region can be described by a few basis functions, while the chemical
  bond is described by the overlapping tails of these atomic orbitals.
  Most techniques in this class are a compromise of, on the one hand,
  a well adapted basis set, where the basis functions are difficult to
  handle, and, on the other hand, numerically convenient basis
  functions such as Gaussians, where the inadequacies are compensated
  by larger basis sets.
\item Pseudopotentials\index{pseudopotential} regard an atom as a
  perturbation of the free electron gas. The most natural basis
  functions for the free electron gas are plane waves. Plane-wave
  basis sets are in principle complete and suitable for sufficiently
  smooth wavefunctions. The disadvantage of the comparably large basis
  sets required is offset by their extreme numerical
  simplicity. Finite plane-wave expansions are, however, absolutely
  inadequate to describe the strong oscillations of the wavefunctions
  near the nucleus.  In the pseudopotential approach the Pauli
  repulsion by the core electrons is therefore described by an
  effective potential that expels the valence electrons from the core
  region.  The resulting wavefunctions are smooth and can be
  represented well by plane waves.  The price to pay is that all
  information on the charge density and wavefunctions near the nucleus
  is lost.
\item Augmented-wave methods\index{augmented-wave method} compose
  their basis functions from atom-like wavefunctions in the atomic
  regions and a set of functions, called envelope functions,
  appropriate for the bonding in between.  Space is divided
  accordingly into atom-centered spheres, defining the atomic regions,
  and an interstitial region in between.  The partial solutions of the
  different regions, are matched with value and derivative at the
  interface between atomic and interstitial regions.
\end{itemize}

The projector augmented-wave method\index{projector augmented-wave
  method} is an extension of augmented wave methods and the
pseudopotential approach, which combines their traditions into a
unified electronic structure method.

After describing the underlying ideas of the various approaches, let us
briefly review the history of augmented wave methods and the
pseudopotential approach. We do not discuss the atomic-orbital based
methods, because our focus is the PAW method and its ancestors.

\section{Augmented wave methods}
The augmented wave methods\index{augmented-wave method} have been
introduced in 1937 by Slater \cite{slater37_pr51_846}. His method was
called augmented plane-wave (APW)\index{APW}\index{augmented plane
  wave method} method. Later Korringa \cite{korringa47_physica13_392},
Kohn and Rostokker \cite{kohn54_pr94_1111} modified the idea, which
lead to the so-called KKR\index{KKR method} method. The basic idea
behind the augmented wave methods has been to consider the electronic
structure as a scattered-electron problem: Consider an electron beam,
represented by a plane wave, traveling through a solid.  It undergoes
multiple scattering at the atoms.  If, for some energy, the outgoing
scattered waves interfere destructively, so that the electrons can not
escape, a bound state has been determined.  This approach can be
translated into a basis-set method with energy- and
potential-dependent basis functions.  In order to make the scattered
wave problem tractable, a model potential had to be chosen: The
so-called muffin-tin potential\index{muffin-tin potential}
approximates the true potential by a potential, that is spherically
symmetric in the atomic regions, and  constant in between.

Augmented-wave methods reached adulthood in the 1970s: O.~K.~Andersen
\cite{andersen75_prb12_3060} showed that the energy dependent basis
set of Slater's APW method can be mapped onto one with energy
independent basis functions by linearizing the partial waves for the
atomic regions with respect to their energy.  In the original APW
approach, one had to determine the zeros of the determinant of an
energy dependent matrix, a nearly intractable numerical problem for
complex systems.  With the new energy independent basis functions,
however, the problem is reduced to the much simpler generalized
eigenvalue problem, which can be solved using efficient numerical
techniques. Furthermore, the introduction of well defined basis sets
paved the way for full-potential calculations
\cite{krakauer79_prb19_1706}. In that case, the muffin-tin
approximation is used solely to define the basis set $|\chi_i\rangle$,
while the matrix elements $\langle\chi_i|H|\chi_j\rangle$ of the
Hamiltonian are evaluated with the full potential.

In the augmented wave methods one constructs the basis set for the
atomic region by solving the radial Schr\"odinger equation for the
spherically averaged effective potential
\begin{eqnarray*}
\left[\frac{-\hbar^2}{2m_e}\vec{\nabla}^2
+v_{eff}(\vec{r})-\epsilon\right]
\phi_{\ell,m}(\epsilon,\vec{r})&=&0
\end{eqnarray*}
as function of the energy. Note that a partial wave
$\phi_{\ell,m}(\epsilon,\vec{r})$ is an angular-momentum eigenstate
and can be expressed as a product of a radial function and a spherical
harmonic. The energy-dependent partial wave is expanded in a Taylor
expansion about some reference energy $\epsilon_{\nu,\ell}$
\begin{eqnarray*}
\phi_{\ell,m}(\epsilon,\vec{r})&=&\phi_{\nu,\ell,m}(\vec{r})
+(\epsilon-\epsilon_{\nu,\ell})\dot\phi_{\nu,\ell,m}(\vec{r})
+O((\epsilon-\epsilon_{\nu,\ell})^2)\;,
\end{eqnarray*}
where
$\phi_{\nu,\ell,m}(\vec{r})=\phi_{\ell,m}(\epsilon_{\nu,\ell},\vec{r})$.
The energy derivative of the partial wave $\dot{\phi}_\nu(\vec{r})
=\left.\frac{\partial\phi(\epsilon,\vec{r})}
  {\partial\epsilon}\right|_{\epsilon_{\nu,\ell}}$ is obtained from
the energy derivative of the Schr\"odinger equation
\begin{eqnarray*}
\left[\frac{-\hbar^2}{2m_e}\vec{\nabla}^2
+v_{eff}(\vec{r})-\epsilon_{\nu,\ell}\right]
\dot\phi_{\nu,\ell,m}(\vec{r})
&=&\phi_{\nu,\ell,m}(\vec{r})\;.
\end{eqnarray*}

Next, one starts from a regular basis set, such as plane waves,
Gaussians or Hankel functions. These basis functions are called
envelope functions $|\tilde\chi_i\rangle$.  Within the atomic region
they are replaced by the partial waves and their energy derivatives,
such that the resulting wavefunction $\chi_i(\vec{r})$ is continuous
and differentiable. The augmented envelope function has the form
\begin{eqnarray}
\chi_i(\vec{r})
=\tilde\chi_i(\vec{r})-\sum_R \theta_R(\vec{r})
\tilde\chi_i(\vec{r})
+\sum_{R,\ell,m}\theta_R(\vec{r})
\left[\phi_{\nu,R,\ell,m}(\vec{r})a_{R,\ell,m,i} 
+\dot\phi_{\nu,R,\ell,m}(\vec{r})b_{R,\ell,m,i} \right]\;.
\label{eq:augmentedwave}
\end{eqnarray}
$\theta_R(\vec{r})$ is a step function that is unity within the
augmentation sphere centered at $\vec{R}_R$ and zero elsewhere.
The augmentation sphere is atom-centered and has a radius about equal
to the covalent radius. This radius is called the muffin-tin radius,
if the spheres of neighboring atoms touch.  These basis functions
describe only the valence states; the core states are localized within
the augmentation sphere and are obtained directly by a radial
integration of the Schr\"odinger equation within the augmentation
sphere.

The coefficients $a_{R,\ell,m,i}$ and $b_{R,\ell,m,i}$ are obtained
for each $|\tilde\chi_i\rangle$ as follows: The envelope function is
decomposed around each atomic site into spherical harmonics multiplied
by radial functions
\begin{eqnarray}
\tilde\chi_i(\vec{r})
=\sum_{\ell,m}u_{R,\ell,m,i}(|\vec{r}-\vec{R}_R|)
Y_{\ell,m}(\vec{r}-\vec{R}_R)\;.
\end{eqnarray}
Analytical expansions for plane waves, Hankel functions or Gaussians exist.
The radial parts of the partial waves $\phi_{\nu,R,\ell,m}$ and
$\dot\phi_{\nu,R,\ell,m}$ are matched with value and derivative to
$u_{R,\ell,m,i}(|\vec{r}|)$, which yields the expansion
coefficients $a_{R,\ell,m,i}$ and $b_{R,\ell,m,i}$.

If the envelope functions are plane waves, the resulting method is
called the linear augmented plane-wave (LAPW) method\index{linear
  augmented-wave method}. If the envelope functions are Hankel
functions, the method is called linear muffin-tin orbital (LMTO)\index{LMTO}
method\index{linear muffin-tin-orbital method}.

A good review of the LAPW method \cite{andersen75_prb12_3060} has been
given by Singh \cite{singh94_book}. Let us now briefly mention the
major developments of the LAPW method: Soler \cite{soler89_prb40_1560}
introduced the idea of additive augmentation: While augmented
plane waves are discontinuous at the surface of the augmentation
sphere if the expansion in spherical harmonics in
Eq.~\ref{eq:augmentedwave} is truncated, Soler replaced the second
term in Eq.~\ref{eq:augmentedwave} by an expansion of the plane wave
with the same angular momentum truncation as in the third term. This
dramatically improved the convergence of the angular momentum
expansion.  Singh \cite{singh91_prb43_6388} introduced so-called local
orbitals, which are non-zero only within a muffin-tin sphere, where
they are superpositions of $\phi$ and $\dot\phi$ functions from
different expansion energies. Local orbitals substantially increase
the energy transferability. Sj\"ostedt \cite{sjoestedt00_ssc113_15}
relaxed the condition that the basis functions are differentiable at
the sphere radius. In addition she introduced local orbitals, which
are confined inside the sphere, and that also have a kink at the
sphere boundary. Due to the large energy cost of kinks, they will
cancel, once the total energy is minimized. The increased variational
degree of freedom in the basis leads to a dramatically improved
plane-wave convergence \cite{madsen01_prb64_195134}.

The second variant of the linear methods is the LMTO method
\cite{andersen75_prb12_3060}.  A good introduction into the LMTO
method is the book by Skriver \cite{skriver84_book}. The LMTO method
uses Hankel functions as envelope functions. The atomic spheres
approximation (ASA)\index{atomic spheres approximation}\index{ASA}
provides a particularly simple and efficient approach to the
electronic structure of very large systems.  In the ASA the
augmentation spheres are blown up so that the sum of their volumes is
equal to the total volume. Then, the first two terms in
Eq.~\ref{eq:augmentedwave} are ignored. The main deficiency of the
LMTO-ASA method is the limitation to structures that can be converted
into a closed packed arrangement of atomic and empty
spheres. Furthermore, energy differences due to structural distortions
are often qualitatively incorrect. Full potential versions of the LMTO
method, that avoid these deficiencies of the ASA have been developed.
The construction of tight binding orbitals as superposition of
muffin-tin orbitals \cite{andersen84_prl53_2571} showed the underlying
principles of the empirical tight-binding method and prepared the
ground for electronic structure methods that scale linearly instead of
with the third power of the number of atoms.  The third generation LMTO
\cite{andersen03_bms26_19} allows to construct true minimal basis
sets, which require only one orbital per electron pair for
insulators. In addition, they can be made arbitrarily accurate in the
valence band region, so that a matrix diagonalization becomes
unnecessary.  The first steps towards a full-potential implementation,
that promises a good accuracy, while maintaining the simplicity 
of the LMTO-ASA method are currently under way.  Through the minimal
basis-set construction the LMTO method offers unrivaled tools for the
analysis of the electronic structure and has been extensively used in
hybrid methods combining density functional theory with model
Hamiltonians for materials with strong electron correlations
\cite{held02_nic10_175}.

\section{Pseudopotentials}
Pseudopotentials\index{pseudopotential} have been introduced to (1)
avoid describing the core electrons explicitly and (2) to avoid the
rapid oscillations of the wavefunction near the nucleus, which
normally require either complicated or large basis sets.

The pseudopotential approach traces back to 1940 when C. Herring
invented the orthogonalized plane-wave method
\cite{herring40_pr57_1169}. Later, Phillips
\cite{phillips59_pr116_287} and Antoncik \cite{antoncik59_jpcs10_314}
replaced the orthogonality condition by an effective potential, which
mimics the Pauli repulsion by the core electrons and thus compensates
the electrostatic attraction by the nucleus. In practice, the
potential was modified, for example, by cutting off the singular
potential of the nucleus at a certain value. This was done with a few
parameters that have been adjusted to reproduce the measured
electronic band structure of the corresponding solid.

Hamann, Schl\"uter and Chiang \cite{hamann79_prl43_1494} showed in
1979 how pseudopotentials can be constructed in such a way, that their
scattering properties are identical to that of an atom to first order
in energy. These first-principles pseudopotentials relieved the
calculations from the restrictions of empirical parameters. Highly
accurate calculations have become possible especially for
semiconductors and simple metals. An alternative approach towards
first-principles pseudopotentials by Zunger and
Cohen\cite{zunger78_prb18_5449} even preceded the one mentioned above.

\subsection*{The idea behind the pseudopotential construction}
In order to construct a first-principles pseudopotential, one starts
out with an all-electron density-functional calculation for a
spherical atom. Such calculations can be performed efficiently on
radial grids.  They yield the atomic potential and wavefunctions
$\phi_{\ell,m}(\vec{r})$.  Due to the spherical symmetry, the
radial parts of the wavefunctions for different magnetic quantum
numbers $m$ are identical.

\begin{figure}[h]
\centering
\includegraphics[width=0.6\textwidth,clip=true]{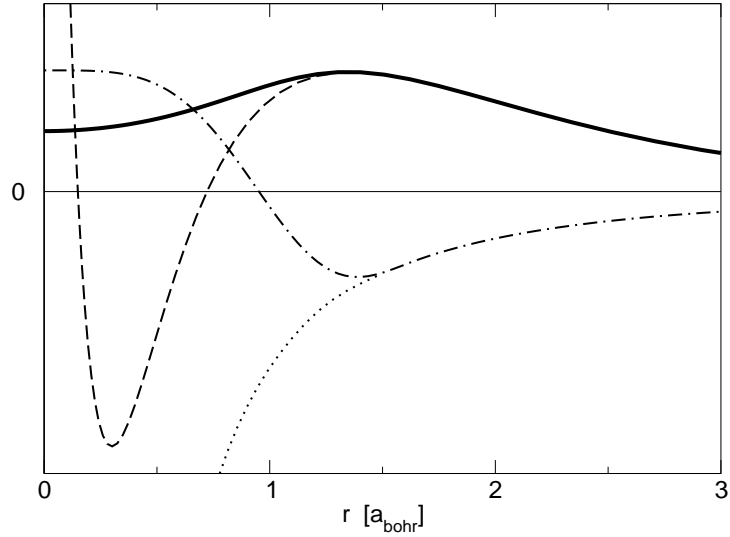}
\caption{Illustration of the pseudopotential concept at the example of
  the 3$s$ wavefunction of Si. The solid line shows the radial part of
  the pseudo wavefunction $\tilde\phi_{\ell,m}$. The dashed line
  corresponds to the all-electron wavefunction $\phi_{\ell,m}$, which
  exhibits strong oscillations at small radii. The angular momentum
  dependent pseudopotential $u_\ell$ (dash-dotted line) deviates from
  the all-electron potential $v_{eff}$ (dotted line) inside the augmentation
  region. The data are generated by the fhi98PP code
  \cite{fuchs98_cpc119_67}.}
\label{fig:pp}
\end{figure}

For the valence wavefunctions one constructs pseudo wavefunctions
$|\tilde\phi_{\ell,m}\rangle$. There are numerous ways
\cite{kerker80_jpcss13_L189, bachelet82_prb26_4199,
  troullier91_prb43_1993, lin93_prb47_4174} to construct those pseudo
wavefunctions: Pseudo wave functions are identical to the true wave
functions outside the augmentation region, which is called core region
in the context of the pseudopotential approach. Inside the
augmentation region the pseudo wavefunction should be node-less and
have the same norm as the true wavefunctions, that is
$\langle\tilde\phi_{\ell,m}|\tilde\phi_{\ell,m}\rangle
=\langle\phi_{\ell,m}|\phi_{\ell,m}\rangle$ (compare
Figure~\ref{fig:pp}).

From the pseudo wavefunction, a potential $u_{\ell}(\vec{r})$ can be
reconstructed by inverting the respective Schr\"odinger equation, i.e.
\begin{eqnarray*}
\left[-\frac{\hbar^2}{2m_e}\vec{\nabla}^2
+u_{\ell}(\vec{r})-\epsilon_{\ell,m}\right]\tilde\phi_{\ell,m}(\vec{r})=0
&\Rightarrow&
u_{\ell}(\vec{r})=\epsilon+\frac{1}{\tilde\phi_{\ell,m}(\vec{r})}
\cdot\frac{\hbar^2}{2m_e}\vec{\nabla}^2\tilde\phi_{\ell,m}(\vec{r})\;.
\end{eqnarray*}
This potential $u_{\ell}(\vec{r})$ (compare Figure~\ref{fig:pp}),
which is also spherically symmetric, differs from one main angular
momentum $\ell$ to the other. Note, that this \textit{inversion of the
  Schr\"odinger equation} works only if the wave functions are
nodeless.

Next we define an effective pseudo Hamiltonian 
\begin{eqnarray}
\hat{\tilde{H}}_\ell=-\frac{\hbar^2}{2m_e}\vec{\nabla}^2+v^{ps}_{\ell}(\vec{r})+
\int
d^3r'\frac{e^2\Bigl(\tilde{n}(\vec{r}')+\tilde{Z}(\vec{r}')\Bigr)}
{4\pi\epsilon_0|\vec{r}-\vec{r'}|} +\mu_{xc}([n(\vec{r})],\vec{r})\;,
\end{eqnarray}
where $\mu_{xc}(\vec{r})=\delta E_{xc}[n]/\delta n(\vec{r})$ is the
functional derivative of the exchange and correlation energy with
respect to the electron density.  Then, we determine the
pseudopotentials $v^{ps}_\ell$ such that the pseudo Hamiltonian
produces the pseudo wavefunctions, that is
\begin{eqnarray}
v^{ps}_{\ell}(\vec{r})&=&u_{\ell}(\vec{r})
-
\int d^3r'\; 
\frac{e^2\Bigl(\tilde{n}(\vec{r}')+\tilde{Z}(\vec{r}')\Bigr)}
{4\pi\epsilon_0|\vec{r}-\vec{r'}|}
-\mu_{xc}([\tilde{n}(\vec{r})],\vec{r})\;.
\label{eq:unscreening}
\end{eqnarray}
This process is called ``unscreening''.\index{unscreening} 

$\tilde{Z}(\vec{r})$ mimics the charge density of the nucleus and
the core electrons. It is usually an atom-centered, spherical Gaussian
that is normalized to the charge of nucleus and core of that atom.  In
the pseudopotential approach, $\tilde{Z}_R(\vec{r})$  does not
change with the potential.  The pseudo density
$\tilde{n}(\vec{r})=\sum_n
f_n\tilde\psi^*_n(\vec{r})\tilde\psi_n(\vec{r})$ is constructed
from the pseudo wavefunctions.

In this way, we obtain a different potential for each angular momentum
channel.  In order to apply these potentials to a given wavefunction,
the wavefunction must first be decomposed into angular momenta.  Then
each component is applied to the pseudopotential $v^{ps}_\ell$ for
the corresponding angular momentum.

The pseudopotential defined in this way can be expressed in a
semi-local form\index{semi-local form}
\begin{eqnarray}
v^{ps}(\vec{r},\vec{r}')
&=&\bar{v}(\vec{r})\delta(\vec{r}-\vec{r}')+
\sum_{\ell,m}
\left[Y_{\ell,m}(\vec{r})
\left[v^{ps}_{\ell}(\vec{r})-\bar{v}(\vec{r})\right] 
\frac{ \delta(|\vec{r}|-|\vec{r}'|)}{|\vec{r}|^2}
Y^*_{\ell,m}(\vec{r}')\right]\;.
\label{eq:semilocal}
\end{eqnarray}
The local potential $\bar{v}(\vec{r})$ only acts on those angular
momentum components that are not already considered explicitely in the
non-local, angular-momentum dependend pseudopotentials $v_\ell^{ps}$.
Typically it is chosen to cancel the most expensive nonlocal terms,
the one corresponding to the highest physically relevant angular
momentum.

The pseudopotential $v^{ps}(\vec{r},\vec{r'})$ is non-local as its
depends on two position arguments, $\vec{r}$ and $\vec{r}'$. The
expectation values are evaluated as a double integral
\begin{eqnarray}
\langle\tilde\psi|\hat{v}_{ps}|\tilde\psi\rangle
&=&\int d^3r\int d^3r'\;\tilde\psi^*(\vec{r})v^{ps}(\vec{r},\vec{r}')
\tilde\psi(\vec{r}')
\end{eqnarray}

The semi-local form of the pseudopotential given in
Eq.~\ref{eq:semilocal} is computationally expensive.  Therefore, in
practice one uses a separable form\index{separable form} of the
pseudopotential \cite{kleinman82_prl48_1425, bloechl90_prb41_5414,
  vanderbilt90_prb41_7892}
\begin{eqnarray}
\hat{v}^{ps}&\approx&\sum_{i,j} \hat{v}^{ps}|\tilde\phi_i\rangle 
\left[\langle\tilde\phi_j|\hat{v}^{ps}|\tilde\phi_i\rangle\right]^{-1}_{i,j} 
\langle\tilde\phi_j|\hat{v}^{ps}\;.
\label{eq:separable}
\end{eqnarray}
Thus, the projection onto spherical harmonics used in the semi-local
form of Eq.~\ref{eq:semilocal} is replaced by a projection onto
angular momentum dependent functions $\hat{v}^{ps}|\tilde\phi_i\rangle$.

The indices $i$ and $j$ are composite indices containing the
atomic-site index $R$, the angular momentum quantum numbers $\ell,m$
and an additional index $\alpha$. The index $\alpha$ distinguishes
partial waves with otherwise identical indices $R,\ell,m$ when more
than one partial wave per site and angular momentum is allowed.  The
partial waves may be constructed as eigenstates to the pseudopotential
$\hat{v}^{ps}_\ell$ for a set of energies.

One can show that the identity of Eq.~\ref{eq:separable} holds by
applying a wavefunction $|\tilde\psi\rangle=\sum_i|\tilde\phi_i\rangle
c_i$ to both sides.  If the set of pseudo partial waves
$|\tilde\phi_i\rangle$ in Eq.~\ref{eq:separable} is complete, the
identity is exact.  The advantage of the separable form is that
$\langle\tilde\phi|\hat{v}^{ps}$ is treated as one function, so that
expectation values are reduced to combinations of simple scalar products
$\langle\tilde\phi_i|\hat{v}^{ps}|\tilde\psi\rangle$.

The total energy of the pseudopotential method can be written in the form
\begin{eqnarray}
E&=&\sum_n f_n
\langle\tilde\psi_n|\frac{\hat{\vec{p}}\,^2}{2m_e}|\tilde\psi_n\rangle
+E_{self}+\sum_n f_n
\langle\tilde\psi_n|\hat{v}_{ps}|\tilde\psi_n\rangle
\nonumber\\
&&+\frac{1}{2} \int d^3r\int d^3r' 
\frac{e^2\Bigl(\tilde{n}(\vec{r})+\tilde{Z}(\vec{r})\Bigr)
\Bigl(\tilde{n}(\vec{r}')+\tilde{Z}(\vec{r}')\Bigr)}
{4\pi\epsilon_0|\vec{r}-\vec{r}'|}
+E_{xc}[\tilde{n}(\vec{r})]\;.
\label{eq:totalenergypseudopotential}
\end{eqnarray}
The constant $E_{self}$ is adjusted such that the total energy of the
atom is the same for an all-electron calculation and the
pseudopotential calculation.

For the atom, from which it has been constructed, this construction
guarantees that the pseudopotential method produces the correct
one-particle energies for the valence states and that the wave
functions have the desired shape. 

While pseudopotentials have proven to be accurate for a large variety
of systems, there is no strict guarantee that they produce the same
results as an all-electron calculation, if they are used in a molecule
or solid. The error sources can be divided into two classes:
\begin{itemize}
\item Energy transferability problems\index{energy transferability}:
  Even for the potential of the reference atom, the scattering
  properties are accurate only in given energy window.
\item Charge transferability problems\index{charge transferability}:
  In a molecule or crystal, the potential differs from that of the
  isolated atom.  The pseudopotential, however, is strictly valid only
  for the isolated atom.
\end{itemize}

The plane-wave basis set for the pseudo wavefunctions is defined by
the shortest wave length $\lambda=2\pi/|\vec{G}|$, where $\vec{G}$ is
the wave vector, via the so-called plane-wave cutoff
$E_{PW}=\frac{\hbar^2G_{max}^2}{2m_e}$ with
$G_{max}=\max\{|\vec{G}|\}$. It is often specified in Rydberg
(1~Ry=$\frac{1}{2}$~H$\approx$13.6~eV).  The plane-wave cutoff is the
highest kinetic energy of all basis functions. The basis-set
convergence can systematically be controlled by increasing the
plane-wave cutoff.

The charge transferability is substantially improved by including a
nonlinear core correction \cite{louie82_prb26_1738} into the
exchange-correlation term of Eq.~\ref{eq:totalenergypseudopotential}.
Hamann \cite{Hamann89} showed, how to construct pseudopotentials also
from unbound wavefunctions. Vanderbilt \cite{Vanderbilt90,Laasonen93}
generalized the pseudopotential method to non-normconserving
pseudopotentials, so-called ultra-soft pseudopotentials, which
dramatically improves the basis-set convergence.  The formulation of
ultra-soft pseudopotentials has already many similarities with the
projector augmented-wave method. Truncated separable pseudopotentials
suffer sometimes from so-called ghost states. These are unphysical
core-like states, which render the pseudopotential useless. These
problems have been discussed by Gonze
\cite{gonze91_prb44_8503}. Quantities such as hyperfine parameters
that depend on the full wavefunctions near the nucleus, can be
extracted approximately \cite{vandewalle93_prb47_4244}. A good review
about pseudopotential methodology has been written by Payne
\cite{Payne92} and Singh \cite{singh94_book}.

In 1985 R. Car and M. Parrinello published the ab-initio molecular
dynamics method \cite{car85_prl55_2471}. Simulations of the atomic
motion have become possible on the basis of state-of-the-art
electronic structure methods. Besides making dynamical phenomena and
finite temperature effects accessible to electronic structure
calculations, the ab-initio molecular dynamics method also introduced
a radically new way of thinking into electronic structure methods.
Diagonalization of a Hamilton matrix has been replaced by classical
equations of motion for the wavefunction coefficients.  If one applies
friction, the system is quenched to the ground state.  Without
friction truly dynamical simulations of the atomic structure are
performed. By using thermostats \cite{nose84_jcp81_511,
  hoover85_pra31_1695, bloechl92_prb45_9413, bloechl02_prb65_104303},
simulations at constant temperature can be performed. The
Car-Parrinello method treats electronic wavefunctions and atomic
positions on an equal footing.
%
\section{Projector augmented-wave method}
\index{projector augmented-wave method} The Car-Parrinello
method\index{Car-Parrinello method} had been implemented first for the
pseudopotential approach. There seemed to be insurmountable barriers
against combining the new technique with augmented wave methods. The
main problem was related to the potential dependent basis set used in
augmented wave methods: the Car-Parrinello method requires a well
defined and unique total energy functional of atomic positions and
basis set coefficients.  Furthermore the analytic evaluation of the
first partial derivatives of the total energy with respect to wave
functions, $\frac{\partial
  E}{\partial\langle\psi_n|}=\hat{H}|\psi_n\rangle f_n$, and atomic
positions, the forces $\vec{F}_j=-\vec{\nabla}_j E$, must be possible.
Therefore, it was one of the main goals of the PAW method to introduce
energy and potential independent basis sets, which were as accurate as
the previously used augmented basis sets. Other requirements have
been: (1) The method should at least match the efficiency of the
pseudopotential approach for Car-Parrinello simulations.  (2) It
should become an exact theory when converged and (3) its convergence
should be easily controlled. We believe that these criteria have been
met, which explains why the PAW method becomes increasingly wide
spread today.

\subsection*{Transformation theory}
At the root of the PAW method lies a transformation, that maps the
true wavefunctions with their complete nodal structure onto auxiliary
wavefunctions, that are numerically convenient.  We aim for smooth
auxiliary wavefunctions, which have a rapidly convergent plane-wave
expansion. With such a transformation we can expand the auxiliary wave
functions into a convenient basis set such as plane waves, and
evaluate all physical properties after reconstructing the related
physical (true) wavefunctions.

Let us denote the physical one-particle wavefunctions as
$|\psi_n\rangle$ and the auxiliary wavefunctions as
$|\tilde\psi_n\rangle$. Note that the tilde refers to the
representation of smooth auxiliary wavefunctions and $n$ is the label
for a one-particle state and contains a band index, a $k$-point and a
spin index. The transformation from the auxiliary to the physical wave
functions is denoted by $\hat{\mathcal{T}}$, i.e.
\begin{eqnarray}
|\psi_n\rangle=\hat{\mathcal{T}}|\tilde{\psi}_n\rangle\;.
\end{eqnarray}

Now we express the constrained density functional $F$ of
Eq.~\ref{eq:ewithconstraint} in terms of our auxiliary wavefunctions
\begin{eqnarray}
F\Bigl[\{\hat{\mathcal{T}}|\tilde\psi_n\rangle\},\{\Lambda_{m,n}\}\Bigr]
= E\Bigl[\{\hat{\mathcal{T}}|\tilde\psi_n\rangle\}\Bigr] 
-\sum_{n,m}\Bigl[\langle\tilde\psi_n|
\hat{\mathcal{T}}^\dagger\hat{\mathcal{T}}
|\tilde\psi_m\rangle -\delta_{n,m}\Bigr]\Lambda_{m,n}\;.
\end{eqnarray}
The variational principle with respect to the auxiliary wavefunctions
yields
\begin{eqnarray}
\hat{\mathcal{T}}^\dagger\hat{H}\hat{\mathcal{T}}|\tilde\psi_n\rangle
=\hat{\mathcal{T}}^\dagger\hat{\mathcal{T}}|\tilde\psi_n\rangle\epsilon_n\,.
\end{eqnarray}
Again, we obtain a Schr\"odinger-like equation (see derivation of
Eq.~\ref{eq:hpsigleichepsi}), but now the Hamilton operator has a
different form, $\hat{\tilde{H}}=\hat{\mathcal{T}}^\dagger
\hat{H}\hat{\mathcal{T}}$, an overlap operator
$\hat{\tilde{O}}=\hat{\mathcal{T}}^\dagger\hat{\mathcal{T}}$ occurs,
and the resulting auxiliary wavefunctions are smooth.

When we evaluate physical quantities, we need to evaluate expectation
values of an operator $\hat{A}$, which can be expressed in terms of
either the true or the auxiliary wavefunctions, i.e.
\begin{eqnarray}
\langle\hat{A}\rangle&=&\sum_nf_n\langle\psi_n|\hat{A}|\psi_n\rangle
=\sum_nf_n\langle\tilde\psi_n|
\hat{\mathcal{T}}^\dagger\hat{A}\hat{\mathcal{T}}|\tilde\psi_n\rangle\,.
\end{eqnarray}
In the representation of auxiliary wavefunctions we need to use
transformed operators
$\hat{\tilde{A}}=\hat{\mathcal{T}}^\dagger\hat{A}\hat{\mathcal{T}}$.
As it is, this equation only holds for the valence electrons.  The
core electrons are treated differently as will be shown below.

The transformation takes us conceptionally from the world of
pseudopotentials to that of augmented wave methods, which deal with
the full wavefunctions. We will see that our auxiliary wavefunctions,
which are simply the plane-wave parts of the full wavefunctions,
translate into the wavefunctions of the pseudopotential approach.  In
the PAW method the auxiliary wavefunctions are used to construct the
true wavefunctions and the total energy functional is evaluated from
the latter.
Thus it provides the missing link between augmented wave methods and the
pseudopotential method, which can be derived as a well-defined approximation
of the PAW method.

In the original paper \cite{bloechl94_prb50_17953}, the auxiliary
wavefunctions have been termed pseudo wavefunctions and the true
wavefunctions have been termed all-electron wavefunctions, in order to
make the connection more evident.  We avoid this notation here,
because it resulted in confusion in cases, where the correspondence is
not clear-cut.
%
\subsection*{Transformation operator}
So far we have described how we can determine the auxiliary wave
functions of the ground state and how to obtain physical information
from them.  What is missing is a definition of the transformation
operator $\hat{\mathcal{T}}$.

The operator $\hat{\mathcal{T}}$ has to modify the smooth auxiliary wave
function in each atomic region, so that the resulting wavefunction
has the correct nodal structure.  Therefore, it makes sense to write
the transformation as identity plus a sum of atomic contributions
$\hat{\mathcal{S}}_R$
\begin{eqnarray}
\hat{\mathcal{T}}=\hat{1}+\sum_R\hat{\mathcal{S}}_R.
\end{eqnarray}
For every atom, $\hat{\mathcal{S}}_R$ adds the difference between the
true and the auxiliary wavefunction. 

The local terms $\hat{\mathcal{S}}_R$ are defined in terms of
solutions $|\phi_{i}\rangle$ of the Schr\"odinger equation for the
isolated atoms.  This set of partial waves $|\phi_{i}\rangle$ will
serve as a basis set so that, near the nucleus, all relevant valence
wavefunctions can be expressed as superposition of the partial waves
with yet unknown coefficients as
\begin{eqnarray}
\psi(\vec{r})=\sum_{i\in R}\phi_{i}(\vec{r}) c_i\quad {\rm
for}\quad |\vec{r}-\vec{R}_R|<r_{c,R}\;.
\end{eqnarray}
With $i\in R$ we indicate those partial waves that belong to site
$R$.

Since the core wavefunctions do not spread out into the neighboring
atoms, we will treat them differently. Currently we use the
frozen-core approximation, which imports the density and the energy of
the core electrons from the corresponding isolated atoms. The
transformation $\hat{\mathcal{T}}$ shall produce only wavefunctions
orthogonal to the core electrons, while the core electrons are treated
separately.  Therefore, the set of atomic partial waves
$|\phi_i\rangle$ includes only valence states that are orthogonal to
the core wavefunctions of the atom.

For each of the partial waves we choose an auxiliary partial wave
$|\tilde\phi_i\rangle$. The identity
\begin{eqnarray}
|\phi_i\rangle&=&(\hat{1}+\hat{\mathcal{S}}_R)|\tilde\phi_i\rangle
\quad\mathrm{for}\quad i\in R 
\nonumber\\
\hat{\mathcal{S}}_R|\tilde\phi_i\rangle&=&|\phi_i\rangle-|\tilde\phi_i\rangle
\label{eq:sr1}
\end{eqnarray}
defines the local contribution $\hat{\mathcal{S}}_R$ to the transformation
operator. Since $\hat{1}+\hat{\mathcal{S}}_R$ shall change the wavefunction only
locally, we require that the partial waves $|\phi_i\rangle$ and their
auxiliary counter parts $|\tilde\phi_i\rangle $ are pairwise identical
beyond a certain radius $r_{c,R}$.
\begin{eqnarray}
\phi_i(\vec{r})
=\tilde\phi_i(\vec{r})
\quad\mathrm{for}\quad i\in R\quad\mathrm{and}\quad|\vec{r}-\vec{R}_R|
>r_{c,R}
\label{eq:equaloutside}
\end{eqnarray}

Note that the partial waves are not necessarily bound states and are
therefore not normalizable, unless we truncate them beyond a certain
radius $r_{c,R}$. The PAW method is formulated such that the final
results do not depend on the location where the partial waves are
truncated, as long as this is not done too close to the nucleus and
identical for auxiliary and all-electron partial waves.

\begin{figure}[ht]
\centering
\includegraphics[width=2.5cm,angle=-90,clip=true]{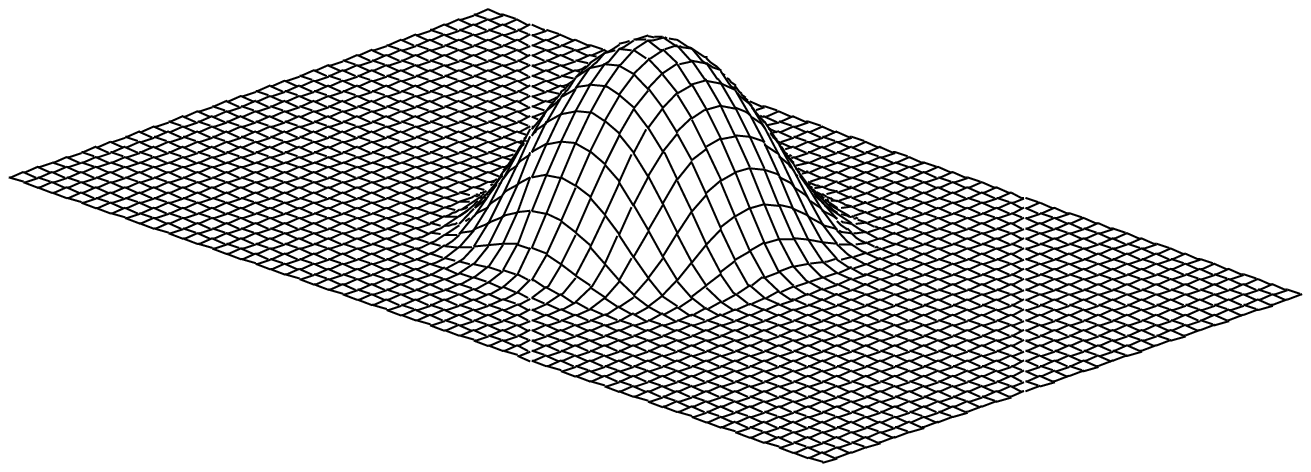}
\includegraphics[width=2.5cm,angle=-90,clip=true]{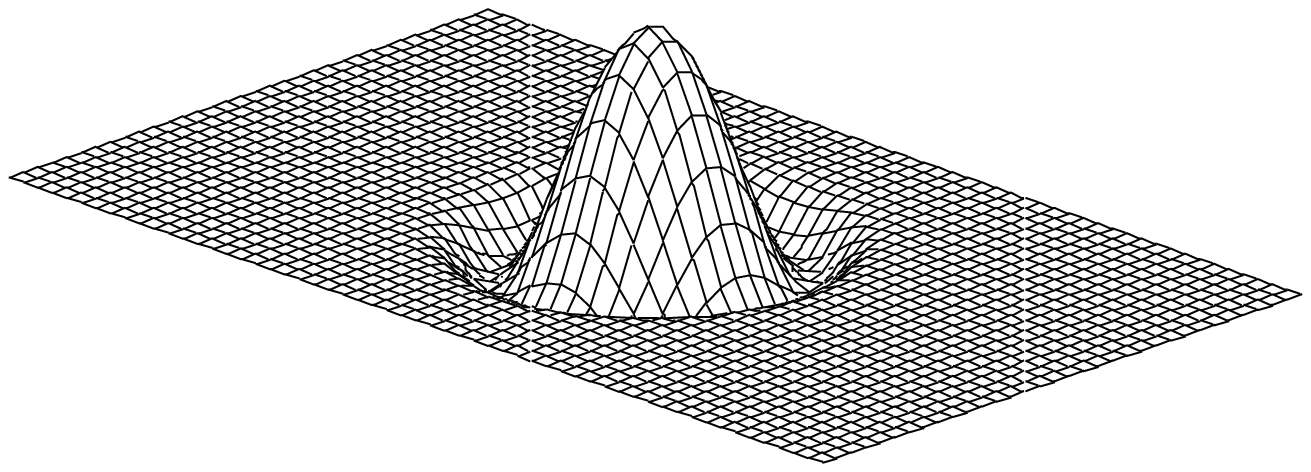}\\
\includegraphics[width=2.5cm,angle=-90,clip=true]{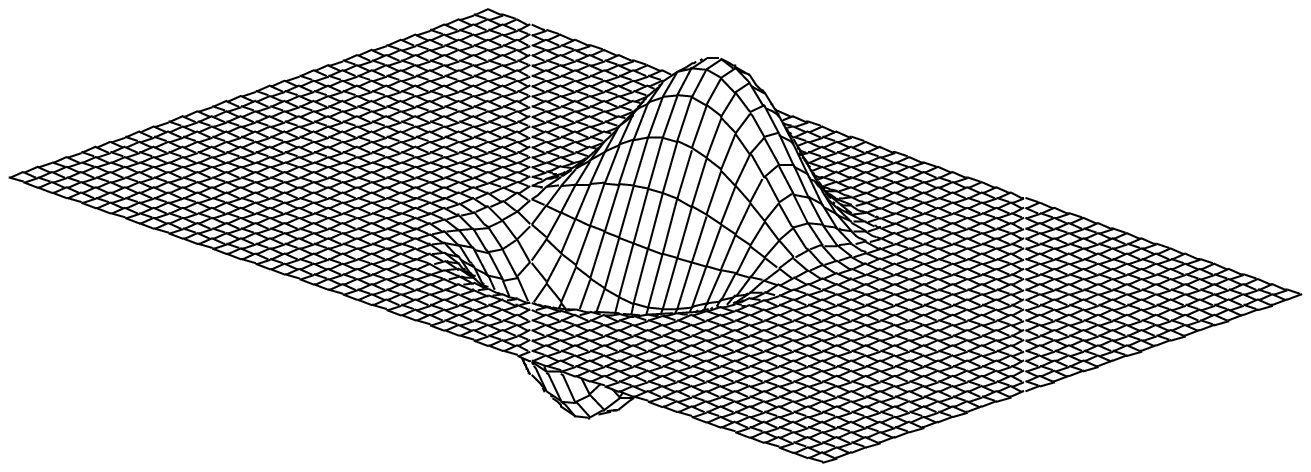}
\includegraphics[width=2.5cm,angle=-90,clip=true]{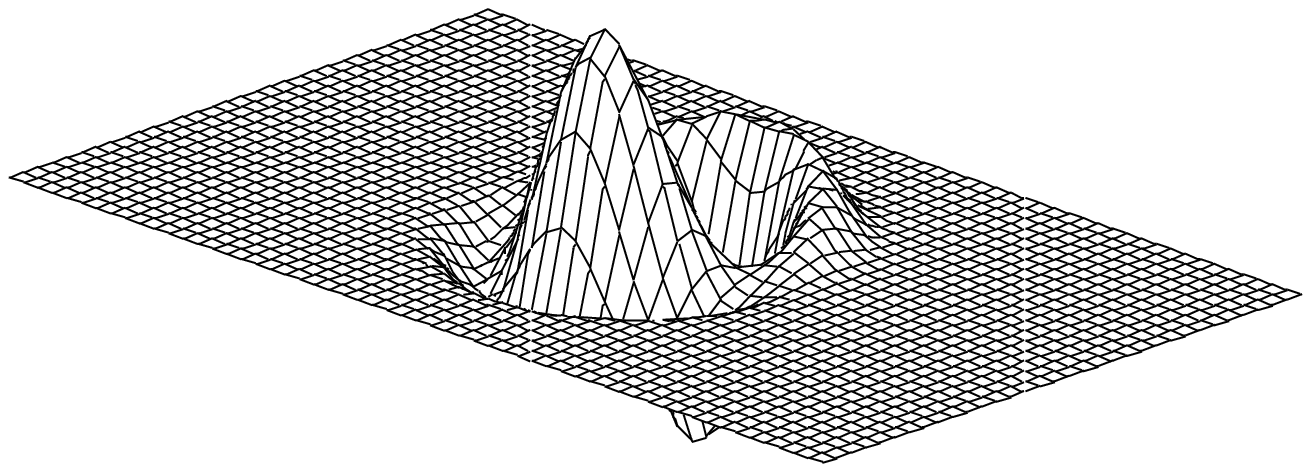}\\
\includegraphics[width=2.5cm,angle=-90,clip=true]{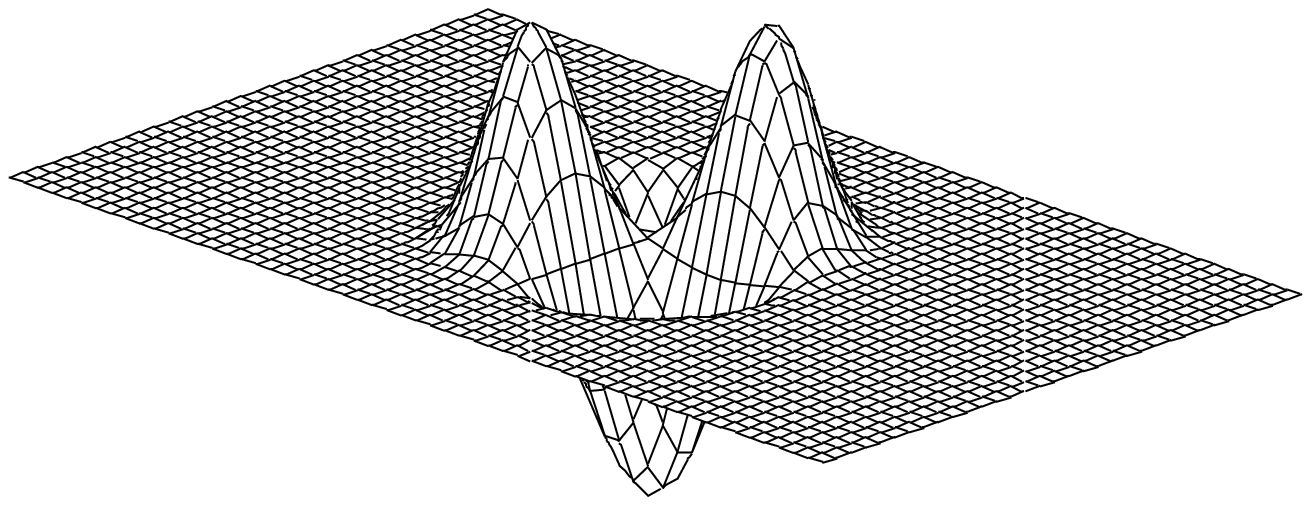}
\caption{Projector functions of the chlorine atom.
Top: two s-type projector functions, middle: p-type, bottom: d-type.}
\label{fig:2}
\end{figure}

In order to be able to apply the transformation operator to an
arbitrary auxiliary wavefunction, we need to be able to expand the
auxiliary wavefunction locally into the auxiliary partial waves
\begin{eqnarray}
\tilde\psi(\vec{r})=\sum_{i\in R} \tilde\phi_i(\vec{r})c_i
=\sum_{i\in R} \tilde\phi_i(\vec{r})
\langle\tilde{p}_i|\tilde\psi\rangle  
\quad{\rm for}\quad |\vec{r}-\vec{R}_R|<r_{c,R}\;,
\label{eq:ps1center}
\end{eqnarray}
which defines the projector functions $|\tilde{p}_i\rangle$.  The
projector functions probe the local character of the auxiliary wave
function in the atomic region.  Examples of projector functions are
shown in Figure~\ref{fig:2}.  From Eq.~\ref{eq:ps1center} we can derive
$\sum_{i\in R}|\tilde\phi_i\rangle\langle\tilde{p}_i|=1$, which is valid
within $r_{c,R}$.  It can be shown by insertion, that the identity
Eq.~\ref{eq:ps1center} holds for any auxiliary wavefunction
$|\tilde\psi\rangle$ that can be expanded locally into auxiliary
partial waves $|\tilde\phi_i\rangle$, if
\begin{equation}
\langle\tilde{p}_i|\tilde\phi_j\rangle=\delta_{i,j} 
\quad\mathrm{for}\quad i,j\in R\,.
\end{equation}
Note that neither the projector functions nor the partial waves need
to be orthogonal among themselves. The projector functions are fully
determined with the above conditions and a closure relation, which is
related to the unscreening of the pseudopotentials (see Eq.~90 in
\cite{bloechl94_prb50_17953}).

By combining Eq.~\ref{eq:sr1} and Eq.~\ref{eq:ps1center}, we can apply
$\hat{\mathcal{S}}_R$ to any auxiliary wavefunction.
\begin{eqnarray}
\hat{\mathcal{S}}_R|\tilde\psi\rangle
&=&\sum_{i\in R} \hat{\mathcal{S}}_R 
|\tilde\phi_i\rangle\langle\tilde{p}_i|\tilde\psi\rangle
=\sum_{i\in R} \Bigl(|\phi_i\rangle-|\tilde\phi_i\rangle\Bigr)
\langle\tilde{p}_i|\tilde\psi\rangle\;.
\end{eqnarray}

Hence, the transformation operator is
\begin{eqnarray}
\hat{\mathcal{T}}=\hat{1}+\sum_i\Bigl(|\phi_i\rangle-|\tilde\phi_i\rangle\Bigr)
\langle\tilde{p}_i|\,,
\label{eq:transf}
\end{eqnarray}
where the sum runs over all partial waves of all atoms.  The true wave
function can be expressed as
\begin{eqnarray}
|\psi\rangle=|\tilde\psi\rangle
+\sum_i\Bigl(|\phi_i\rangle-|\tilde\phi_i\rangle\Bigr)
\langle\tilde{p}_i|\tilde\psi\rangle
=|\tilde\psi\rangle
+\sum_R\Bigl( |\psi^1_R\rangle-|\tilde\psi^1_R\rangle\Bigr)
\label{eq:aewave}
\end{eqnarray}
with
\begin{eqnarray}
|\psi^1_R\rangle&=&\sum_{i\in R}|\phi_i\rangle
\langle\tilde{p}_i|\tilde\psi\rangle
\\
|\tilde\psi^1_R\rangle&=&\sum_{i\in R}|\tilde\phi_i\rangle
\langle\tilde{p}_i|\tilde\psi\rangle\,.
\end{eqnarray}

\begin{figure}[h]
\centering
\includegraphics[height=10cm,angle=-90,clip=true]{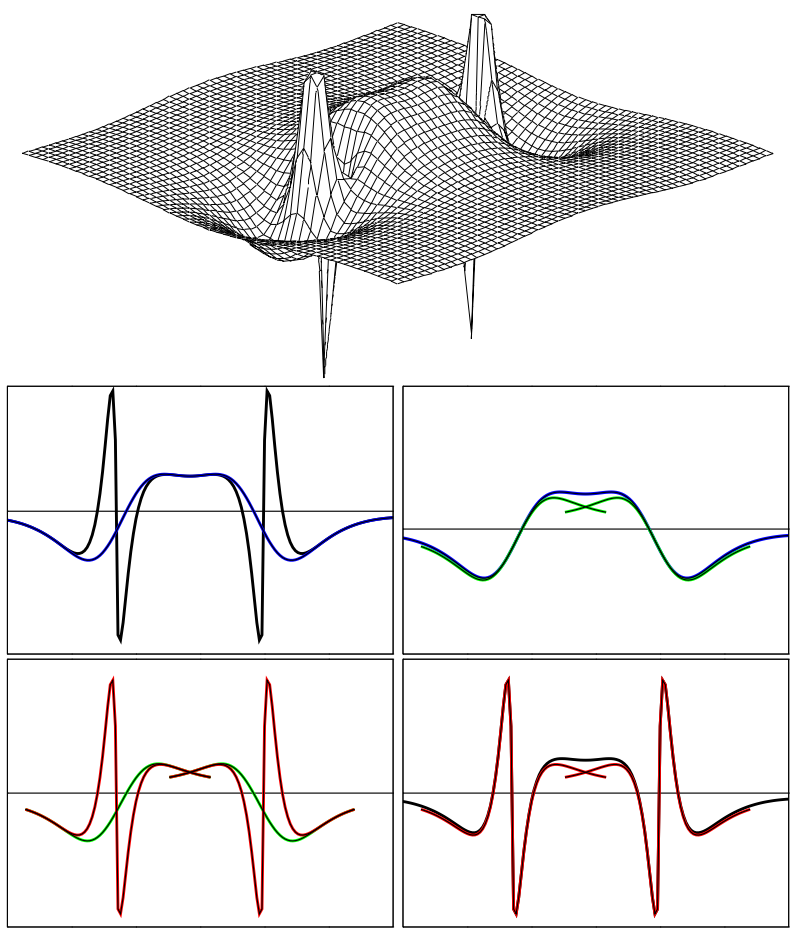}\\
\includegraphics[width=5.1cm,clip=true]{ae-ps_new.eps}
\includegraphics[width=5.1cm,clip=true]{ps-ps1_new.eps}\\
\includegraphics[width=5.1cm,clip=true]{ae1-ps1_new.eps}
\includegraphics[width=5.1cm,clip=true]{ae-ae1_new.eps}
\caption{Bonding p-$\sigma$ orbital of the Cl$_2$ molecule and its
decomposition of the wavefunction into auxiliary wavefunction and the
two one-center expansions.  Top-left: True and auxiliary wave
function; top-right: auxiliary wavefunction and its partial wave
expansion; bottom-left: the two partial wave expansions; bottom-right:
true wavefunction and its partial wave expansion.}
\label{fig:1}
\end{figure}

In Fig.~\ref{fig:1} the decomposition of Eq.~\ref{eq:aewave} is shown
for the example of the bonding p-$\sigma$ state of the Cl$_2$
molecule.

To understand the expression Eq.~\ref{eq:aewave} for the true wave
function, let us concentrate on different regions in space. (1) Far
from the atoms, the partial waves are, according to
Eq.~\ref{eq:equaloutside}, pairwise identical so that the auxiliary
wavefunction is identical to the true wavefunction, that is
$\psi(\vec{r})=\tilde\psi(\vec{r})$.  (2) Close to an atom $R$,
however, the auxiliary wavefunction is, according to
Eq.~\ref{eq:ps1center}, identical to its one-center expansion, that is
$\tilde\psi(\vec{r})=\tilde\psi^1_R(\vec{r})$.  Hence the true
wavefunction $\psi(\vec{r})$ is identical to
$\psi^1_R(\vec{r})$, which is built up from partial waves that
contain the proper nodal structure.

In practice, the partial wave expansions are truncated.  Therefore,
the identity of Eq.~\ref{eq:ps1center} does not hold strictly.  As a
result, the plane waves also contribute to the true wavefunction
inside the atomic region. This has the advantage that the missing
terms in a truncated partial wave expansion are partly accounted for
by plane waves. This explains the rapid convergence of the
partial wave expansions. This idea is related to the additive
augmentation of the LAPW method of Soler \cite{soler89_prb40_1560}.

Frequently, the question comes up, whether the transformation
Eq.~\ref{eq:transf} of the auxiliary wavefunctions indeed provides
the true wavefunction. The transformation should be considered merely
as a change of representation analogous to a coordinate transform.  If
the total energy functional is transformed consistently, its minimum
will yield auxiliary wavefunctions that produce the correct wave
functions $|\psi\rangle$.

%
\subsection*{Expectation values}
Expectation values can be obtained either from the reconstructed
true wavefunctions or directly from the auxiliary wave
functions 
\begin{eqnarray}
\langle\hat{A}\rangle&=&\sum_{n}f_n\langle\psi_n|\hat{A}|\psi_n\rangle
+\sum_{n=1}^{N_c}\langle\phi_n^c|\hat{A}|\phi_n^c\rangle
\nonumber\\
&=&\sum_{n}f_n\langle\tilde\psi_n|\hat{\mathcal{T}}^\dagger \hat{A}\hat{\mathcal{T}}
|\tilde\psi_n\rangle
+\sum_{n=1}^{N_c}\langle\phi_n^c|\hat{A}|\phi_n^c\rangle\,,
\label{eq:expect1}
\end{eqnarray}
where $f_n$ are the occupations of the valence states and $N_c$ is the
number of core states. The first sum runs over the valence states, and
second over the core states $|\phi^c_n\rangle$.

Now we can decompose the matrix element for a wavefunction $\psi$ into
its individual contributions according to Eq.~\ref{eq:aewave}.
\begin{eqnarray}
\langle\psi|\hat{A}|\psi\rangle&=&
\langle\tilde\psi+\sum_R(\psi^1_R-\tilde\psi^1_R)|
\hat{A}|\tilde\psi+\sum_{R'}(\psi^1_{R'}-\tilde\psi^1_{R'})\rangle
\nonumber\\
&=&\underbrace{\langle\tilde\psi|\hat{A}|\tilde\psi\rangle
+\sum_R\Bigl(\langle\psi^1_R|\hat{A}|\psi^1_R\rangle
-\langle\tilde\psi^1_R|\hat{A}|\tilde\psi^1_R\rangle\Bigr)}_{\mbox{part 1}}
\nonumber\\
&+&\underbrace{\sum_R\Bigl(
\langle \psi^1_R-\tilde\psi^1_R|\hat{A}|\tilde\psi-\tilde\psi^1_R\rangle
+\langle\tilde\psi-\tilde\psi^1_R|\hat{A}|\psi^1_R-\tilde\psi^1_R\rangle
\Bigr)}_{\mbox{part 2}}
\nonumber\\
&+&\underbrace{\sum_{R\neq R'}\langle \psi^1_R
-\tilde\psi^1_R|\hat{A}|\psi^1_{R'}-\tilde\psi^1_{R'}\rangle}
_{\mbox{part 3}}
\label{eq:expect}
\end{eqnarray}
Only the first part of Eq.~\ref{eq:expect}, is evaluated explicitly,
while the second and third parts of Eq.~\ref{eq:expect} are neglected,
because they vanish for sufficiently local operators as long as the
partial wave expansion is converged: The function
$\psi^1_R-\tilde\psi^1_R$ vanishes per construction beyond its
augmentation region, because the partial waves are pairwise identical
beyond that region. The function $\tilde\psi-\tilde\psi^1_R$ vanishes
inside its augmentation region, if the partial wave expansion is
sufficiently converged.  In no region of space both functions
$\psi^1_R-\tilde\psi^1_R$ and $\tilde\psi -\tilde\psi^1_R$ are
simultaneously nonzero.  Similarly the functions
$\psi^1_R-\tilde\psi^1_R$ from different sites are never non-zero in
the same region in space. Hence, the second and third parts of
Eq.~\ref{eq:expect} vanish for operators such as the kinetic energy
$\frac{-\hbar^2}{2m_e}\vec{\nabla}^2$ and the real space projection
operator $|r\rangle\langle r|$, which produces the electron density.
For truly nonlocal operators the parts 2 and 3 of Eq.~\ref{eq:expect}
would have to be considered explicitly.

The expression, Eq.~\ref{eq:expect1}, for the expectation value can
therefore be written with the help of Eq.~\ref{eq:expect} as
\begin{eqnarray}
\langle\hat{A}\rangle&=&
\sum_{n}f_n\Bigl(\langle\tilde\psi_n|\hat{A}|\tilde\psi_n\rangle
+\langle\psi^1_n|\hat{A}|\psi^1_n\rangle
-\langle\tilde\psi^1_n|\hat{A}|\tilde\psi^1_n\rangle\Bigr)
+\sum_{n=1}^{N_c}\langle\phi^c_n|\hat{A}|\phi^c_n\rangle
\nonumber\\
&=&\sum_{n}f_n\langle\tilde\psi_n|\hat{A}|\tilde\psi_n\rangle
+\sum_{n=1}^{N_c}\langle\tilde\phi^c_n|\hat{A}|\tilde\phi^c_n\rangle
\nonumber\\
&+&\sum_R\Bigl(\sum_{i,j\in R}D_{i,j}\langle\phi_j|\hat{A}|\phi_i\rangle
+\sum_{n\in R}^{N_{c,R}}\langle\phi^c_n|\hat{A}|\phi^c_n\rangle\Bigr)
\nonumber\\
&-&\sum_R\Bigl(\sum_{i,j\in R} D_{i,j}\langle\tilde\phi_j|\hat{A}|\tilde\phi_i\rangle
+\sum_{n\in R}^{N_{c,R}}\langle\tilde\phi^c_n|\hat{A}|\tilde\phi^c_n\rangle\Bigr)
\,,
\end{eqnarray}
where $\mat{D}$ is the one-center density matrix defined as
\begin{eqnarray}
D_{i,j}=\sum_n
f_n\langle\tilde\psi_n|\tilde{p}_j\rangle
\langle\tilde{p}_i|\tilde\psi_n\rangle
=\sum_n \langle\tilde{p}_i|\tilde\psi_n\rangle
f_n\langle\tilde\psi_n|\tilde{p}_j\rangle\,.
\label{eq:1cdenmat}
\end{eqnarray}

The auxiliary core states, $|\tilde\phi^c_n\rangle$ allow to
incorporate the tails of the core wavefunction into the plane-wave
part, and therefore assure, that the integrations of partial wave
contributions cancel each other strictly beyond $r_c$. They are
identical to the true core states in the tails, but are a smooth
continuation inside the atomic sphere. It is not required that the
auxiliary wave functions are normalized.

Following this scheme, the electron density is given by
\begin{eqnarray}
n(\vec{r})&=&\tilde{n}(\vec{r})
+\sum_R\Bigl(n^1_R(\vec{r})-\tilde{n}^1_R(\vec{r})\Bigr)
\\
\tilde{n}(\vec{r})
&=&\sum_n f_n \tilde\psi^*_n(\vec{r})\tilde\psi_n(\vec{r})+\tilde{n}_c(\vec{r})
\nonumber\\
n^1_R(\vec{r})&=&\sum_{i,j\in R}
D_{i,j}\phi^*_j(\vec{r})\phi_i(\vec{r})+n_{c,R}(\vec{r})
 \nonumber
\nonumber\\
\tilde{n}^1_R(\vec{r})&=&\sum_{i,j\in R}
D_{i,j}\tilde\phi^*_j(\vec{r})\tilde\phi_i(\vec{r})+\tilde{n}_{c,R}(\vec{r})
\,,
\end{eqnarray}
where $n_{c,R}$ is the core density of the corresponding atom and
$\tilde{n}_{c,R}$ is the auxiliary core density that is identical to $n_{c,R}$
outside the atomic region, but smooth inside.

Before we continue, let us discuss a special point: The matrix elements
of a general operator with the auxiliary wavefunctions may be slowly
converging with the plane-wave expansion, because the operator
$\hat{A}$ may not be well behaved. An example for such an operator is
the singular electrostatic potential of a nucleus. This problem can be
alleviated by adding an ``intelligent zero'': If an operator $\hat{B}$
is purely localized within an atomic region, we can use the identity
between the auxiliary wavefunction and its own partial wave expansion
\begin{eqnarray}
0&=&\langle\tilde\psi_n|\hat{B}|\tilde\psi_n\rangle
-\langle\tilde\psi_n^1|\hat{B}|\tilde\psi_n^1\rangle\,.
\label{eq:zeroop}
\end{eqnarray}

Now we choose an operator $\hat{B}$ so that it cancels the problematic
behavior of the operator $\hat{A}$, but is localized in a single
atomic region.  By adding $\hat{B}$ to the plane-wave part and the
matrix elements with its one-center expansions, the plane-wave
convergence can be improved without affecting the converged result. A
term of this type, namely $\hat{\bar{v}}$ will be introduced in the next
section to cancel the Coulomb singularity of the potential at the
nucleus.

\subsection*{Total energy}
Like wavefunctions and expectation values, also the total
energy can be divided into three parts.
\begin{eqnarray}
E\Bigl[\{|\tilde\psi_n\rangle\},\{R_R\}\Bigr]
&=&\tilde{E}+\sum_R\Bigl(E^1_R-\tilde{E}^1_R\Bigr)
\end{eqnarray}
The plane-wave part $\tilde{E}$ involves only smooth functions and is
evaluated on equi-spaced grids in real and reciprocal space.  This part is
computationally most demanding, and is similar to the expressions in the
pseudopotential approach.
\begin{eqnarray}
\tilde{E}&=&\sum_n\langle\tilde\psi_n|\frac{\hat{\vec{p}}\,^2}{2m_e}
|\tilde\psi_n\rangle
+\frac{1}{2}
\int d^3r\int d^3r'\;
\frac{e^2\Bigl(\tilde{n}(\vec{r})+\tilde{Z}(\vec{r})\Bigr)
\Bigl(\tilde{n}(\vec{r}')+\tilde{Z}(\vec{r}')\Bigr)}
{4\pi\epsilon_0|\vec{r}-\vec{r}'|}
\nonumber\\
&+&\int d^3r\; \bar{v}(\vec{r})\tilde{n}(\vec{r})
+E_{xc}[\tilde{n}]
\label{eq:psetot}
\end{eqnarray}
$\tilde{Z}({\bf r})$ is an angular-momentum dependent core-like
density that will be described in detail below.  The remaining parts
can be evaluated on radial grids in a spherical-harmonics expansion.
The nodal structure of the wavefunctions can be properly described on
a logarithmic radial grid that becomes very fine near the nucleus,
\begin{eqnarray}
{E}^1_R&=&\sum_{i,j\in R} D_{i,j}
\langle\phi_j|\frac{\hat{\vec{p}}\,^2}{2m_e}|\phi_i\rangle
+\sum_{n\in R}^{N_{c,R}}
\langle\phi^c_n|\frac{\hat{\vec{p}}\,^2}{2m_e}|\phi^c_n\rangle
\nonumber\\
&+&\frac{1}{2}
\int d^3r\int d^3r'\;
\frac{e^2\Bigl(n^1(\vec{r})+Z(\vec{r})\Bigr)
\Bigl(n^1(\vec{r'})+Z(\vec{r'})\Bigr)}
{|\vec{r}-\vec{r}'|}
+E_{xc}[n^1]
\\
\tilde{E}^1_R&=&\sum_{i,j\in R} D_{i,j}
\langle\tilde\phi_j|\frac{\hat{\vec{p}}\,^2}{2m_e}|\tilde\phi_i\rangle
+\frac{1}{2}
\int d^3r\int d^3r'\;
\frac{e^2\Bigl(\tilde{n}^1(\vec{r})+\tilde{Z}(\vec{r})\Bigr)
\Bigl(\tilde{n}^1(\vec{r'})+\tilde{Z}(\vec{r'})\Bigr)}
{4\pi\epsilon_0|\vec{r}-\vec{r}'|}
\nonumber\\
&+&
\int d^3r\; \bar{v}(\vec{r})\tilde{n}^1(\vec{r})
+E_{xc}[\tilde{n}^1]\,.
\label{eq:ps1etot}
\end{eqnarray}

The compensation charge density
$\tilde{Z}(\vec{r})=\sum_R\tilde{Z}_R(\vec{r})$ is given as a
sum of angular momentum dependent Gauss functions, which have an
analytical plane-wave expansion.  A similar term occurs also in the
pseudopotential approach. In contrast to the norm-conserving
pseudopotential approach, however, the compensation charge of an atom
$\tilde{Z}_R$ is non-spherical and constantly adapts 
instantaneously to the environment. It is constructed such that
\begin{eqnarray}
n^1_R(\vec{r})+Z_R(\vec{r}) -\tilde{n}^1_R(\vec{r})
-\tilde{Z}_R(\vec{r})
\end{eqnarray}
has vanishing electrostatic multi-pole moments for each atomic site.
With this choice, the electrostatic potentials of the augmentation
densities vanish outside their spheres. This is the reason that there is
no electrostatic interaction of the one-center parts between different
sites.

The compensation charge density as given here is still localized
within the atomic regions. A technique similar to an Ewald summation,
however, allows to replace it by a very extended charge density. Thus
we can achieve, that the plane-wave convergence of the total energy
is not affected by the auxiliary density.

The potential $\bar{v}=\sum_R\bar{v}_R$, which occurs in
Eqs.~\ref{eq:psetot} and~\ref{eq:ps1etot} enters the total energy in
the form of ``intelligent zeros'' described in Eq.~\ref{eq:zeroop}
\begin{eqnarray}
0=\sum_nf_n\left(\langle\tilde\psi_n|\bar{v}_R|\tilde\psi_n\rangle
-\langle\tilde\psi^1_n|\bar{v}_R|\tilde\psi^1_n\rangle\right)
=\sum_nf_n\langle\tilde\psi_n|\bar{v}_R|\tilde\psi_n\rangle
-\sum_{i,j\in R}D_{i,j}
\langle\tilde\phi_i|\bar{v}_R|\tilde\phi_j\rangle\,.
\end{eqnarray}
The main reason for introducing this potential is to cancel the
Coulomb singularity of the potential in the plane-wave part. The
potential $\bar{v}$ allows to influence the plane-wave convergence
beneficially, without changing the converged result.  $\bar{v}$ must
be localized within the augmentation region, where
Eq.~\ref{eq:ps1center} holds.
%
\subsection*{Approximations}
Once the total energy functional provided in the previous section has
been defined, everything else follows: Forces are partial derivatives
with respect to atomic positions.  The potential is the derivative of
the non-kinetic energy contributions to the total energy with respect
to the density, and the auxiliary Hamiltonian follows from derivatives
$\tilde{H}|\tilde\psi_n\rangle$ with respect to auxiliary wave
functions.  The fictitious-Lagrangian approach of Car and Parrinello
\cite{Car85} does not allow any freedom in the way these derivatives
are obtained. Anything else than analytic derivatives will violate
energy conservation in a dynamical simulation. Since the expressions
are straightforward, even though rather involved, we will not discuss
them here.

All approximations are incorporated already in the total energy functional of
the PAW method. What are those approximations?
\begin{itemize}
\item Firstly we use the frozen-core approximation. In principle this
approximation can be overcome.
\item The plane-wave expansion for the auxiliary wavefunctions must
be complete. The plane-wave expansion is controlled easily by
increasing the plane-wave cutoff defined as
$E_{PW}=\frac{1}{2}\hbar^2G_{max}^2$.  Typically we use a plane-wave
cutoff of 30~Ry.
\item The partial wave expansions must be converged. Typically we use
one or two partial waves per angular momentum $(\ell,m)$ and site.
It should be noted that the partial wave expansion is not
variational, because it changes the total energy functional and not
the basis set for the auxiliary wavefunctions.
\end{itemize}
Here, we do not discuss here numerical approximations such as the choice of
the radial grid, since those are easily controlled.

\subsection*{Relation to pseudopotentials}
We mentioned earlier that the pseudopotential approach can be derived
as a well defined approximation from the PAW method: The
augmentation part of the total energy $\Delta E=E^1-\tilde{E}^1$ for
one atom is a functional of the one-center density matrix $\mat{D}$
defined in Eq.~\ref{eq:1cdenmat}. The pseudopotential approach can be
recovered if we truncate a Taylor expansion of $\Delta E$ about the
atomic density matrix after the linear term.  The term linear in
$\mat{D}$ is the energy related to the nonlocal pseudopotential.
\begin{eqnarray}
\Delta E(\mat{D})&=&\Delta E(\mat{D}^{at})
+\sum_{i,j}
\left.\frac{\partial \Delta E}{\partial D_{i,j}}\right|_{\mat{D}^{at}}
(D_{i,j}-D^{at}_{i,j})
+O(\mat{D}-\mat{D}^{at})^2
\nonumber\\
&=&E_{self}+\sum_n f_n\langle\tilde\psi_n|\hat{v}^{ps}|\tilde\psi_n\rangle
-\int d^3r\; \bar{v}(\vec{r})\tilde{n}(\vec{r})
+O(\mat{D}-\mat{D})^2\,,
\end{eqnarray}
which can directly be compared to the total energy expression
Eq.~\ref{eq:totalenergypseudopotential} of the pseudopotential
method. The local potential $\bar{v}(\vec{r})$ of the pseudopotential
approach is identical to the corresponding potential of the projector
augmented-wave method. The remaining contributions in the PAW total
energy, namely $\tilde{E}$, differ from the corresponding terms in
Eq.~\ref{eq:totalenergypseudopotential} only in two features: our
auxiliary density also contains an auxiliary core density, reflecting
the nonlinear core correction of the pseudopotential approach, and the
compensation density $\tilde{Z}(\vec{r})$ is non-spherical and depends
on the wave function.  Thus we can look at the PAW method also as a
pseudopotential method with a pseudopotential that adapts
instantaneously to the electronic environment. In the PAW method, the
explicit nonlinear dependence of the total energy on the one-center
density matrix is properly taken into account.

What are the main advantages of the PAW method compared to the
pseudopotential approach?  

Firstly all errors can be systematically controlled so that there are
no transferability errors.  As shown by Watson \cite{Watson98} and
Kresse \cite{kresse99_prb59_1758}, most pseudopotentials fail for high
spin atoms such as Cr. While it is probably true that pseudopotentials
can be constructed that cope even with this situation, a failure can
not be known beforehand, so that some empiricism remains in practice:
A pseudopotential constructed from an isolated atom is not guaranteed
to be accurate for a molecule.  In contrast, the converged results of
the PAW method do not depend on a reference system such as an isolated
atom, because PAW uses the full density and potential.

Like other all-electron methods, the PAW method provides access to the
full charge and spin density, which is relevant, for example, for
hyperfine parameters. Hyperfine parameters are sensitive probes of the
electron density near the nucleus. In many situations they are the only
information available that allows to deduce atomic structure and
chemical environment of an atom from experiment.

The plane-wave convergence is more rapid than in norm-conserving
pseudopotentials and should in principle be equivalent to that of
ultra-soft pseudo\-potentials \cite{Vanderbilt90}. Compared to the
ultra-soft pseudo\-potentials, however, the PAW method has the
advantage that the total energy expression is less complex and can
therefore be expected to be more efficient.

The construction of pseudopotentials requires to determine a number of
parameters. As they influence the results, their choice is critical.
Also the PAW methods provides some flexibility in the choice of
auxiliary partial waves. However, this choice does not influence the
converged results.

\subsection*{Recent developments}
Since the first implementation of the PAW method in the CP-PAW code
\cite{bloechl94_prb50_17953}, a number of groups have adopted the PAW
method.  The second implementation was done by the group of Holzwarth
\cite{Holzwarth97}. The resulting PWPAW code is freely available
\cite{Holzwarth01}.  This code is also used as a basis for the PAW
implementation in the ABINIT project \cite{torrent08_cms42_337}. An
independent PAW code has been developed by Valiev and Weare
\cite{Valiev99}. This implementation has entereed the NWChem code
\cite{windus03_lncs2660_168}. An independent implementation of the PAW
method is that of the VASP code \cite{kresse99_prb59_1758}. The PAW
method has also been implemented by W. Kromen \cite{kromen01_thesis}
into the EStCoMPP code of Bl\"ugel and Schr\"oder. Another
implementation is in the Quantum Espresso code
\cite{gianozzi09_jpcm21_395502}.  A real-space-grid based version of
the PAW method is the code GPAW developed by Mortensen et
al. \cite{mortensen05_prb71_35109}.

Another branch of methods uses the reconstruction of the PAW method,
without taking into account the full wavefunctions in the energy
minimization. Following chemist notation, this approach could be termed
``post-pseudopotential PAW''. This development began with the
evaluation for hyperfine parameters from a pseudopotential calculation
using the PAW reconstruction operator \cite{Hyperfine} and is now
used in the pseudopotential approach to calculate properties that
require the correct wavefunctions such as hyperfine parameters.

The implementation of the PAW method by Kresse and Joubert
\cite{kresse99_prb59_1758} has been particularly useful as they had an
implementation of PAW in the same code as the ultra-soft
pseudo\-potentials, so that they could critically compare the two
approaches with each other. Their conclusion is that both methods
compare well in most cases, but they found that magnetic energies are
seriously -- by a factor two -- in error in the pseudo\-potential
approach, while the results of the PAW method were in line with other
all-electron calculations using the linear augmented plane-wave
method. As a short note, Kresse and Joubert incorrectly claim that
their implementation is superior as it includes a term that is
analogous to the non-linear core correction of pseudo\-potentials
\cite{Louie82}: this term however is already included in the original
version in the form of the pseudized core density.

Several extensions of the PAW have been done in the recent years:
For applications in chemistry truly isolated systems are often of
great interest. As any plane-wave based method introduces periodic
images, the electrostatic interaction between these images can cause
serious errors. The problem has been solved by mapping the charge
density onto a point charge model, so that the electrostatic
interaction could be subtracted out in a self-consistent
manner  \cite{decoupling}. In order to include the influence of the
environment, the latter was simulated by simpler force fields using
the molecular-mechanics-quantum-mechanics (QM-MM) approach  \cite{QMMM}.

In order to overcome the limitations of the density functional theory
several extensions have been performed. Bengone \cite{Bengone00}
implemented the LDA+U approach into our CP-PAW code.  Soon after this,
Arnaud \cite{Arnaud2000} accomplished the implementation of the GW
approximation into our CP-PAW code.  The VASP-version of PAW
\cite{Kresse00} and our CP-PAW code have now been extended to include
a non-collinear description of the magnetic moments. In a
non-collinear description, the Schr\"odinger equation is replaced by
the Pauli equation with two-component spinor wavefunctions.

The PAW method has proven useful to evaluate electric field
gradients  \cite{EFG} and magnetic hyperfine parameters with high
accuracy  \cite{SiO2}. Invaluable will be the prediction of NMR chemical
shifts using the GIPAW method of Pickard and Mauri  \cite{Mauri2001},
which is based on their earlier work  \cite{Mauri96}. While the GIPAW
is implemented in a post-pseudo\-potential manner, the extension to a
self-consistent PAW calculation should be straightforward.  An
post-pseudo\-potential approach has also been used to evaluate core
level spectra  \cite{Pickard97} and momentum matrix
elements  \cite{Kageshima}.

\subsection*{Acknowledgement}
Part of this article has been written together with Clemens F\"orst
and Johannes K\"astner. I am also grateful for the careful reading and
helpful suggestions by Rolf Fader, Johannes Kirschner, J\"urgen Noffke
and Philipp Seichter.  Financial support by the Deutsche
Forschungsgemeinschaft through FOR 1346 is gratefully acknowledged.


\section*{Appendices}
\appendix
\section{Model  exchange-correlation energy}
\label{app:modelxc}

We consider a model with a constant density, and a hole function,
which describes a situation, where all electrons of the same spin are
repelled completely from a sphere centered at the reference electron

The hole function has the form
\begin{eqnarray*}
h(\vec{r},\vec{r}_0)=
\begin{cases}
-\frac{1}{2}n(\vec{r}_0)&\text{for $|\vec{r}-\vec{r}_0|<r_h$}\\
0 & \text{otherwise}\\
\end{cases}
\end{eqnarray*}
where $n(\vec{r})$ is the electron density and the hole radius
$r_h=\sqrt[3]{\frac{2}{4\pi n}}$ is the radius of the sphere, which is
determined such that the exchange correlation hole integrates to $-1$,
i.e. $\frac{4\pi}{3}r_h^3 \bigl(\frac{1}{2}n\bigr)=1$.

The potential of a homogeneously charged sphere with
radius $r_h$ and one positive charge is 
\begin{eqnarray*}
v(r)=\frac{e^2}{4\pi\epsilon_0}
\begin{cases}
-\frac{3}{2r_h}+\frac{1}{2r_h}\left(\frac{r}{r_h}\right)^2
&\text{for $r\le r_h$}\\
-\frac{1}{r}&\text{for $r>r_h$}\\
\end{cases}
\end{eqnarray*}
where $r=|\vec{r}-\vec{r}_0|$.

With Eq.~\ref{eq:uxc} we obtain for the potential contribution of the
exchange correlation energy 
\begin{eqnarray*}
U_{xc}=-\int d^3r\; n(\vec{r}) v(r=0)
=-\int d^3r\; \frac{e^2}{4\pi\epsilon_0} 
\frac{3}{4}\sqrt[3]{\frac{2\pi}{3}}\cdot n^{\frac{4}{3}}
\end{eqnarray*}

\section{Large-gradient limit of the enhancement factor}
\label{app:fofx}

An exponentially decaying density 
\begin{eqnarray}
n(r)=\exp(-\lambda r)
\label{eq:app2nofr}
\end{eqnarray}
has a reduced gradient
\begin{eqnarray}
x:=\frac{|\vec{\nabla}n|}{n^{\frac{4}{3}}}= \lambda\exp(+\frac{1}{3}\lambda r)
\label{eq:app2xofr}
\end{eqnarray}

We make the following ansatz for the exchange correlation energy per
electron
\begin{eqnarray}
\epsilon_{xc}(n,x)=-Cn^{\frac{1}{3}}F(x)
\end{eqnarray}
where only the local exchange has been used and $C$ is a constant.

Enforcing the long-distance limit of the exchange correlation energy 
per electron for exponentially decaying densities
\begin{eqnarray}
\epsilon_{xc}((n(r),x(r))=-\frac{1}{2}\frac{e^2}{4\pi\epsilon_0 r}
\end{eqnarray}
yields
\begin{eqnarray}
F(x)=\frac{e^2}{4\pi\epsilon_0 r(x) 2C n^\frac{1}{3}(r(x))}
\end{eqnarray}
Using Eqs.~\ref{eq:app2nofr} and \ref{eq:app2xofr}, we express the
radius and the density by the reduced gradient, i.e.
\begin{eqnarray}
r(x)&=&-\frac{3}{\lambda}\biggl(\ln[\lambda]-\ln[x]\biggr)
\\
n(x)&=&n(r(x))=\lambda^3 x^{-3}\;,
\end{eqnarray}
and obtain
\begin{eqnarray}
F(x)&=&\frac{e^2}{4\pi\epsilon_0 
\biggl[-\frac{3}{\lambda}\biggl(\ln[\lambda]-\ln[x]\biggr)\biggr]
\biggl[2C \lambda x^{-1}\biggr]}
=\bigl(\frac{e^2}{4\pi\epsilon_0\cdot 6C}\bigr)
\frac{x^2}{x\ln(\lambda)-x\ln(x)}
\nonumber\\
&\stackrel{x\rightarrow\infty}{\rightarrow}&
-\biggl(\frac{e^2}{4\pi\epsilon_0\cdot 6C}\biggr)\frac{x^2}{x\ln(x)}
\end{eqnarray}
Now we need to ensure that $F(0)=1$ so that the gradient correction
vanishes for the homogeneous electron gas, and that $F(x)=F(-x)$ to
enforce spin reversal symmetry. There are several possible
interpolations for these requirements, but the most simple one is
\begin{eqnarray}
F(x)=1-\frac{\beta x^2}{1+\frac{4\pi\epsilon_0}{e^2}\cdot 6C\beta x\cdot\mathrm{asinh}(x)}
\end{eqnarray}
This is the enhancement factor for exchange used by Becke in his B88
functional \cite{becke88_pra38_3098}.

\end{document}

%% file: fig2a.tex
\resizebox{0.9\linewidth}{!}{\setlength{\unitlength}{3cm}
  \begin{picture}(3.5,2.)(-1,-2.5)\large\sf
     \linethickness{0.5mm}
     \renewcommand{\xscale}{20}
     \renewcommand{\yscale}{0.1}
      \put(0.0,-2.3){{\Huge 0.0}}
      \put(1.85,-2.3){{\Huge 0.1}}
      \put(3.7,-2.3){{\Huge 0.2}}
      \put(2.7,-2.3){{\Huge Na}}
      \put(-0.7,-0.05){{\Huge 0~eV}}
      \put(-1.0,-1.05){{\Huge -10~eV}}
      \put(-1.0,-2.05){{\Huge -20~eV}}
      \put(2.0,-2.6){{\Huge $n$}}
      \put(0,-2.){\framebox(4,2){}}
      \scaleput(0.07,-5){{\Huge $\varepsilon_{xc}$}}
      \scaleput(0.125,-9){{\Large LSD}}    \scaleput(0,0){\curve(0.12,-9,0.09846,-10.726)}
      \scaleput(0.175,-10){{\Large HF}}    \scaleput(0,0){\curve(0.17,-10,0.14329,-10.516)}
      \scaleput(0.02,-18){{\Large model}} \scaleput(0,0){\curve(0.06,-16,0.08007,-11.254)}
\curvedashes[2mm]{0,1,2}
\scaleput(0,0){\curve(0.00010, -0.933,0.00013, -1.016,0.00033, -1.393,0.00089, -1.931,0.00197, -2.517,0.00374, -3.120,0.00640, -3.730,0.01010, -4.344,0.01502, -4.959,0.02135, -5.575,0.02924, -6.192,0.03889, -6.809,0.05046, -7.426,0.06413, -8.044,0.08007, -8.662,0.09846, -9.280,0.11947, -9.898,0.14329,-10.516,0.17007,-11.135,0.20000,-11.753)}
\curvedashes{}
\scaleput(0,0){\curve(0.00010, -1.350,0.00013, -1.457,0.00033, -1.936,0.00089, -2.594,0.00197, -3.289,0.00374, -3.988,0.00640, -4.682,0.01010, -5.371,0.01502, -6.054,0.02135, -6.732,0.02924, -7.406,0.03889, -8.076,0.05046, -8.742,0.06413, -9.406,0.08007,-10.067,0.09846,-10.726,0.11947,-11.383,0.14329,-12.038,0.17007,-12.691,0.20000,-13.342)}
\curvedashes[2mm]{0,1,2}
\scaleput(0,0){\curve(0.00010, -1.212,0.00013, -1.320,0.00033, -1.810,0.00089, -2.509,0.00197, -3.271,0.00374, -4.054,0.00640, -4.847,0.01010, -5.643,0.01502, -6.443,0.02135, -7.243,0.02924, -8.045,0.03889, -8.847,0.05046, -9.649,0.06413,-10.451,0.08007,-11.254,0.09846,-12.057,0.11947,-12.860,0.14329,-13.663,0.17007,-14.467,0.20000,-15.270)}
   \end{picture}
  }